\documentclass[journal,10pt]{IEEEtran}
\usepackage{graphicx}
\usepackage{subfigure}
\usepackage{float}
\usepackage{subfloat}
\usepackage{amsmath}
\usepackage{amssymb}
\usepackage{amsthm}
\usepackage{epstopdf}
\usepackage{cases}
\usepackage{color}
\usepackage{makecell}
\usepackage{cite}
\usepackage{algorithmic}
\usepackage[ruled]{algorithm2e}

\usepackage[normalem]{ulem}

\ifCLASSINFOpdf
\else
\fi
\usepackage{caption}
\usepackage{mathtools}

\begin{document}
\setlength{\textfloatsep}{5pt}

%
\title{Complex CNN CSI Enhancer for Integrated Sensing and Communications}
%
%
%

\author{Xu Chen,~\IEEEmembership{Member,~IEEE,}
	Zhiyong Feng,~\IEEEmembership{Senior Member,~IEEE,}
	J. Andrew Zhang,~\IEEEmembership{Senior Member,~IEEE,}\\
	Feifei Gao,~\IEEEmembership{Fellow,~IEEE,}
	Xin Yuan,~\IEEEmembership{Member,~IEEE,}
	Zhaohui Yang,~\IEEEmembership{Member,~IEEE,}\\
	and Ping Zhang,~\IEEEmembership{Fellow,~IEEE}
	\thanks{This work has been submitted to the IEEE for possible publication. Copyright may be transferred without notice, after which this version may no longer be accessible.}
	\thanks{Xu Chen, Z. Feng are with School of Information and communication Engineering, Beijing University of Posts and Telecommunications, Beijing 100876, P. R. China (Email:\{chenxu96330, fengzy\}@bupt.edu.cn).}
	\thanks{J. A. Zhang is with the Global Big Data Technologies Centre, University of Technology Sydney, Sydney, NSW 2007, Australia (Email: Andrew.Zhang@uts.edu.au).}
	\thanks{F. Gao is with Institute for Artificial Intelligence, Tsinghua University (THUAI), State Key Lab of Intelligent Technologies and Systems, Tsinghua University, Beijing National Research Center for Information Science and Technology (BNRist), Department of Automation, Tsinghua University, Beijing 100084, P. R. China (Email: feifeigao@ieee.org).}
	\thanks{X. Yuan is with the Data61, Commonwealth Scientific and Industrial Research Organization, Sydney, NSW 2122, Australia (Email: xin.yuan@data61.csiro.au).}
	\thanks{Z. Yang is with the College of Information Science and Electronic Engineering, Zhejiang University, Hangzhou 310007, China, and Zhejiang Provincial Key Lab of Information Processing, Communication and Networking (IPCAN), Hangzhou 310007, P. R. China (Email: yang\_zhaohui@zju.edu.cn).}
	\thanks{Ping Zhang is with State Key Laboratory of Networking and Switching Technology, Beijing University of Posts and Telecommunications, Beijing 100876, P. R. China (Email: pzhang@bupt.edu.cn).}
	\thanks{Corresponding author: Zhiyong Feng.}
}

%
%

\markboth{}%
{Shell \MakeLowercase{\textit{et al.}}: Bare Demo of IEEEtran.cls for IEEE Journals}
%


\maketitle

\newcounter{mytempeqncnt}
\setcounter{mytempeqncnt}{\value{equation}}
\begin{abstract}
In this paper, we propose a novel complex convolutional neural network (CNN) CSI enhancer for integrated sensing and communications (ISAC), which exploits the correlation between the sensing parameters (such as angle-of-arrival and range) and the channel state information (CSI) to significantly improve the CSI estimation accuracy and further enhance the sensing accuracy. Within the CNN CSI enhancer, we use the complex-valued computation layers to form the CNN, which maintains the phase information of CSI. We also transform the CSI into the sparse angle-delay domain, leading to heatmap images with prominent peaks that can be efficiently processed by CNN. Based on the enhanced CSI outputs, we further propose a novel biased fast Fourier transform (FFT)-based sensing scheme for improving the range sensing accuracy, by artificially introducing phase biasing terms. Extensive simulation results show that the ISAC complex CNN CSI enhancer can converge within 30 training epochs. The normalized mean square error (NMSE) of its CSI estimates is about 17 dB lower than that of the linear minimum mean square error (LMMSE) estimator, and the bit error rate (BER) of demodulation using the enhanced CSI estimation approaches that with perfect CSI. Finally, the range estimation MSE of the proposed biased FFT-based sensing method approaches that of the subspace-based sensing method, at a much lower complexity.

\end{abstract}


%
\IEEEpeerreviewmaketitle

\section{Introduction}
%
%
%
%

\subsection{Background and Motivations}
Integrated sensing and communications (ISAC), also known as joint communications and sensing (JCAS), has been regarded as one of the most promising techniques for improving the spectrum efficiency in the future sixth generation (6G) networks~\cite{liu2020joint, Saad2020, Feng2021JCSC}. 
ISAC achieves wireless communication and sensing using the same transmit signals~\cite{Zhang2022ISAC, Valkama2022, IBFDJCR}. It can estimate the sensing parameters such as angle-of-arrival (AoA), delay (or range), and Doppler frequency shift from the channel state information (CSI) between the base station (BS) and the user equipment (UE)~\cite{ZhangOverviewJCS, Feng2021, 2023XuJCAS}. 
Therefore, the CSI estimation is crucial for both the reliability of communication demodulation and the sensing accuracy, and there is a strong correlation between the sensing parameters and CSI.

The least-square (LS) method has been widely used for CSI estimation due to its low complexity~\cite{2010MIMO}. However, its low CSI estimation accuracy, especially in the low signal-to-noise ratio (SNR) regime, may cause severe deterioration in communication reliability. The minimum mean square error (MMSE) method was proposed to improve the CSI estimation accuracy. However, the high complexity makes it challenging to be utilized in real applications~\cite{Rodger2014principles}. In this paper, by treating the CSIs as image tensors, we intend to use the complex convolutional neural network (CNN) combined with the ISAC transforms to enhance the CSI estimation for improving both the communication reliability and sensing accuracy of the ISAC system.

\subsection{Related Works}

In \cite{2023ChenKalman}, the authors proposed a Kalman filter-based CSI enhancer for the ISAC system. It exploited the sensing parameters estimated by ISAC sensing schemes as the prior information for constructing the state transfer of the CSI, and a Kalman filter used the reconstructed state transfer to suppress the noise terms in the initially estimated CSI. Since the Kalman filter has a recurrent optimization structure, it can be regarded as a simple neural network (NN). Further, there have been many studies using more complicated deep neural networks (DNN) to achieve the CSI denoiser to improve the CSI estimation accuracy~\cite{Yuwei2022, Valkama2021, 2018YeDL, 2019MehrabiDNNCE,2020MaCE}. In \cite{2018YeDL}, the authors proposed that end-to-end deep fully connected networks can outperform the conventional MMSE estimator under insufficient pilots. In \cite{2019MehrabiDNNCE}, the authors proposed to use DNN to enhance the decision-directed channel estimation for time-varying channels. Moreover, the DNN channel estimator can be jointly designed with beamforming and precoding for massive multiple-input and multiple-output (MIMO) channels~\cite{2020MaCE}.

Recently, it has been demonstrated that CNN has enormous potential for exploiting the correlation between the CSI elements and the time, frequency, and spatial domain features~\cite{Ottersten2022,Kumar2022,2019DLCESoltani,2019DLCEDong,2021JiangdualCNN}. In~\cite{2019DLCESoltani}, the authors showed that the CNN-based CSI denoiser outperformed the traditional MMSE method. In~\cite{2019DLCEDong}, the authors proposed to exploit the spatial-frequency-domain CSI to obtain better CSI estimation performance than the MMSE method at a lower complexity. Considering that the sparsity of CSI in the transformed domain is a remarkable feature for improving CSI estimation~\cite{2018GaoCS}, the authors in \cite{2021JiangdualCNN} proposed a dualCNN structure to improve the CSI estimation accuracy.

There are two major issues with the aforementioned CNN CSI enhancers. Firstly, they are designed by treating the real and imaginary parts of the complex-valued CSI as two isolated real-valued channels to learn the complex-valued operations of channel response and signal transmission, which is inefficient in keeping the phase information of the complex-valued CSI. Secondly, according to the existing ISAC works~\cite{2023XuJCAS,2021YangJEVAR,2018GaoCS}, the CSIs in the transformed angle-delay domain are usually much more sparse in expression, and the sensing parameters, such as angles, delays, etc., are correlated to the CSI. As a classic CSI enhancement method, the transform-domain channel estimation technique~\cite{ZhuTransformedCSI} truncates CSI in the transform-domain around the peak values and pads zero values to the truncated transform-domain CSI to suppress the normalized mean squared error (NMSE) of CSI estimation. However, this results in the removal of a majority of tails of the CSI transform and thereby disrupts the phase relations between CSI elements. Therefore, this method is destructive for sensing based on the phase terms of CSI and causes a larger deviation of CSI estimation compared with the MMSE method.

\subsection{Our Contributions}
In this paper, we aim to address the above two major issues of CSI enhancers. We propose a novel ISAC complex CNN CSI enhancer for efficiently suppressing the noise in the CSI estimates with a relatively low complexity, which provides highly accurate CSI estimation for both reliable communication demodulation and accurate sensing estimation. We also propose a novel biased fast Fourier transform (FFT)-based sensing method to improve the range estimation accuracy with complexity close to the FFT-based sensing method.

Since the radio channel response is complex-valued, instead of using DNNs with two real-valued channels to train the CSI denoiser by treating the imaginary and real parts of complex CSI as two isolated real values, we use the complex-valued computation layers to construct a complex CNN CSI enhancer, which well maintains the phase information of the complex channel response. Moreover, we add FFT-based ISAC transform modules into the CNN to convert the CSI from the antenna-time domain to the sparse angle-delay domain, which is similar to the radar heatmap images and can be more efficiently processed by CNN~\cite{Sturm2011, 2021JiangdualCNN}. Since the parameter sensing is based on the CSI, we set minimizing the NMSE of CSI estimation as the objective for the training of ISAC complex CNN to unify communication and sensing optimization. Simulation results show that the training descending speed of the complex CNN CSI enhancer is fast, and the NMSE loss of the CSI enhancer is remarkably low. 

The main contributions of this paper are summarized as follows.
\begin{itemize}
	\item[1.] We propose an ISAC complex CNN CSI enhancer that uses the complex-valued computation layers to maintain the phase information of complex-valued signal transmission instead of treating the imaginary and real parts of CSI as two isolated real values. Furthermore, we integrate the FFT-based ISAC transform modules into the complex CNN structure, which can transform the CSI from the original domain into the sparse angle-delay domain. The CSI in the delay-angle domain is similar to the heatmap image, which can be well-processed by CNN to extract useful feature maps. The loss descending speed and generalization performance of CNN benefit from these settings.
	
	\item[2.] Since the accuracy of AoA estimation based on the enhanced CSI estimates is significantly improved, we introduce a spatial filter to separate the CSIs of individual channel paths according to their AoA estimates, then estimate the propagation delay for each path and demodulate the communication signals simultaneously.
	
	\item[3.] We propose a biased FFT-based sensing method for improved range estimation. This method first actively adds known biasing phase shift terms to the CSI estimates to generate multiple biased sensing results via the FFT-based sensing, and then obtain the improved estimation via the mean of all debiased results. We prove that the average sensing error of the proposed method is always lower than that achieved by using the FFT-based sensing method.
\end{itemize}

We provide extensive simulation results, which verify that the CSI estimation accuracy can be remarkably improved by the proposed ISAC complex CNN enhancer such that the bit error rate (BER) of communication demodulation and the accuracy of sensing can also be improved significantly.

The remaining parts of this paper are organized as follows. 
In Section \ref{sec:system-model}, we describe the system model and basic methods in the ISAC system. 
Section \ref{sec:ISAC_CNN_processing} introduces the ISAC complex CNN CSI enhancer in detail.
Section \ref{sec:ISAC_processing} introduces the AoA and range estimation method based on the enhanced CSI estimates.
In Section \ref{sec:Simulation}, the simulation results are presented. 
Section \ref{sec:conclusion} concludes this paper.

\begin{figure}[!t]
	\centering
	\includegraphics[width=0.20\textheight]{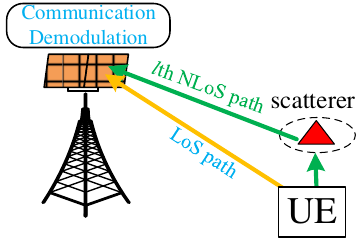}%
	\DeclareGraphicsExtensions.
	\caption{The ISAC scenario.}
	\label{fig: Uplink JCS Model}
\end{figure}

\textbf{Notations}: Bold uppercase letters denote matrices (e.g., $\textbf{M}$); bold lowercase letters denote column vectors (e.g., $\textbf{v}$); scalars are denoted by normal font (e.g., $\gamma$); the entries of vectors or matrices are referred to with brackets, for instance, the $q$th entry of vector $\textbf{v}$ is $[\textbf{v}]_{q}$, and the entry of the matrix $\textbf{M}$ at the $m$th row and $q$th column is ${[\textbf{M}]_{n,m}}$; ${{\bf{U}}_s} = {\left[ {\bf{U}} \right]_{:,{N_1}:{N_2}}}$ means the matrices sliced from the $N_1$th to the $N_2$th columns of $\bf U$; $\left(\cdot\right)^{*}$ and $\left(\cdot\right)^T$ denote Hermitian transpose, complex conjugate and transpose, respectively; ${\bf M}_1 \in \mathbb{C}^{M\times N}$ and ${\bf M}_2 \in \mathbb{R}^{M\times N}$ are ${M\times N}$ complex-valued and real-valued matrices, respectively; ${\left\| {\bf{H}} \right\|_F}$ is the Frobenius norm of a matrix $\bf H$; for two given matrices ${\bf{S}}1$ and ${\bf{S}}2$,  ${ {[ {{v_{p,q}}} ]} |_{(p,q) \in {\bf{S}}1 \times {\bf{S}}2}}$ denotes the vector stacked by values ${v_{p,q}}$ satisfying $p\in{\bf{S}}1$ and $q\in{\bf{S}}2$; and $v \sim \mathcal{CN}(m,\sigma^2)$ means $v$ follows a circular symmetric complex Gaussian (CSCG) distribution with mean $m$ and variance $\sigma^2$.

\section{System Model}\label{sec:system-model}

This section presents the ISAC system setup, MIMO channel model, brief introduction of conventional channel estimators, and FFT-based ISAC transform to provide fundamentals for illustrating the complex CNN CSI enhancer.

\subsection{ISAC System Setup}
Due to the reciprocity of the time-division-duplex (TDD) system, we consider the uplink (UL) channel estimation for the ISAC system, where the BS is equipped with a uniform linear array (ULA), and the user equipment (UE) has one single antenna, as shown in Fig.~\ref{fig: Uplink JCS Model}. In the UL preamble (ULP) period, BS uses the received training sequences in the preamble signals transmitted by UE for CSI estimation, and the estimated CSI is further used to estimate sensing parameters, such as the AoA and range of UE. In the UL data (ULD) period, the BS demodulates the UL data signals of UE using the estimated CSI. Orthogonal frequency division multiplexing (OFDM) signal is adopted as the transmit signal. For simplicity of presentation, we assume that there is one ULP in each packet for CSI estimation, and the ULPs are transmitted at an equal interval, denoted by $T_s^p$. Moreover, synchronization between the BS and UEs is assumed to be achieved via a global clock, such as GPS disciplined oscillator (GPSDO). Thus, there are no clock asynchronism issues, such as the timing and frequency offsets, or sensing ambiguity, as discussed in \cite{Zhang2022ISAC}.

The critical parameters for the OFDM signal are represented as follows. ${P_t^U}$ is the transmit power, ${N_c}$ is the number of subcarriers occupied by UE; ${M_s}$ is the number of OFDM packets used for each sensing parameter estimation; ${d_{n,m}}$ is the transmit OFDM baseband symbol of the $m$th OFDM symbol at the $n$th subcarrier, $f_c$ is the carrier frequency, $\Delta f$ is the subcarrier interval, ${T_s}$ is the time duration of each OFDM symbol, and $T_s^p = P_s T_s$, where $P_s$ is the number of OFDM symbols in each packet. The sensing parameter estimates are updated every ${M_s}$ packets.


\subsection{Channel Model} \label{subsec:channel_model}
This subsection presents the multi-path channel model for the MIMO-OFDM system. The uniform interval between neighboring antenna elements is denoted by $d_a$. The size of ULA is ${P} \times {1}$. The AoA for receiving or the angle-of-departure (AoD) for transmitting the $k$th far-field signal is ${\theta _k}$. The steering vector of the ULA is~\cite{HAARDT2014651}
\begin{equation}\label{equ:phase_difference}
	{\bf{a}}({\theta _k}) = {[ {1,{e^{j\frac{{2\pi }}{\lambda }d_a\sin {\theta _k}}},\cdots, {e^{j\frac{{2\pi }}{\lambda }(P - 1)d_a\sin {\theta _k}}}} ]^T} \in \mathbb{C}^{P \times 1},
\end{equation}
where $\lambda = c/f_c$ is the wavelength of the carrier, and $c$ is the speed of light. 

Further, the multi-path channel model for the $n$th subcarrier of the $m$th packet can be expressed as 
\begin{equation}\label{equ:channel_vec}
	{{\bf{h}}_{n,m}}{\rm{ = }}\sum\limits_{l = 0}^{L - 1} {\left[ {{b_{C,l}}{e^{j2\pi m{T_s^p}{f_{d,l}}}}{e^{ - j2\pi n\Delta f{\tau _l}}}{\bf{a}}({\theta _l})} \right]}  \in {\mathbb{C}^{P \times 1}},
\end{equation}
where $L$ is the number of propagation paths, $l = 0$ is for the channel response of the line-of-sight (LoS) path, and $l \in \{1, \cdots, L-1\}$ is for the paths involved with the $l$th scatterer; ${\bf{a}}({\theta _l})$ is the steering vector for UL receiving and transmission, respectively; $\theta _l$ is the corresponding AoA; ${f_{d,0}} = \frac{{{v_{0}}}}{\lambda }$ and ${\tau _{0}} = \frac{{{r_{0,1}}}}{c}$ are the Doppler shift and delay between UE and BS of the line-of-sight (LoS) path, respectively, with ${v_{0}}$ and ${r_{0,1}}$ being the corresponding radial relative velocity and the distance, respectively; ${f_{d,l}} = {f_{d,l,1}} + {f_{d,l,2}}$ and ${\tau _{l}} = {\tau _{l,1}} + {\tau _{l,2}}$ are the aggregate Doppler shift and delay of the $l$th non-line-of-sight (NLoS) path, respectively; ${f_{d,l,1}} = \frac{{{v_{l,1}}}}{\lambda }$ and ${f_{d,l,2}} = \frac{{{v_{l,2}}}}{\lambda }$ are the Doppler shifts between UE and the $l$th scatterer, and between the $l$th scatterer and BS, respectively, with ${v_{l,1}}$ and ${v_{l,2}}$ being the corresponding radial velocities; ${\tau _{l,1}} = \frac{{{r_{l,1}}}}{c}$ and ${\tau _{l,2}} = \frac{{{r_{l,2}}}}{c}$ are the delays between UE and the $l$th scatterer, and between BS and the $l$th scatterer, respectively, with ${r_{l,1}}$ and ${r_{l,2}}$ being the corresponding distances. Moreover, ${b_{C,0}} = \sqrt {\frac{{{\lambda ^2}}}{{{{(4\pi {r_{0,1}})}^2}}}}$ and ${b_{C,l}} = \sqrt {\frac{{{\lambda ^2}}}{{{{\left( {4\pi } \right)}^3}{r_{l,1}}^2{r_{l,2}}^2}}} {\beta _{C,l}}$ are the attenuation of the LoS and NLoS paths, respectively; and ${\beta _{C,l}}$ is the reflecting factor of the $l$th scatterer, following $\mathcal{CN}(0,\sigma _{C\beta ,l}^2)$~\cite{Rodger2014principles}.

\subsection{Channel Estimator}
The training sequences of the $n$th subcarrier of the $m$th packet transmitted by UE are received by BS, and the received signal is expressed as
\begin{equation}\label{equ:receive_vec}
	{{\bf{Y}}_{n,m}} = \sqrt {{P_t}} {{\bf{h}}_{n,m}} \otimes {\left( {{\bf{s}}_{n,m}^u} \right)^H} + {{\bf{Z}}_{n,m}},
\end{equation}
where ${P_t}$ is the transmit power, $ \otimes $ is the Kronecker product, and ${{\bf{Z}}_{n,m}}$ is the Gaussian noise matrix with each element following independent and identically distributed (i.i.d.) $\mathcal{CN}(0,\sigma _n^2)$. Moreover, ${\bf{s}}_{n,m}^u \in {\mathbb{C}^{U \times 1}}$ is one of the orthogonal codebook, which satisfies
\begin{equation}\label{equ:orthogonal_codebook}
	{\left( {{\bf{s}}_{n,m}^{{u_1}}} \right)^H}{\bf{s}}_{n,m}^{{u_2}} = \left\{ \begin{array}{l}
		U,{u_1} = {u_2},\\
		0,{u_1} \ne {u_2}.
	\end{array} \right.
\end{equation}

The LS method is widely used due to its low complexity. The LS estimate of CSI is given by
\begin{equation}\label{equ:LS_est}
	{\bf{\hat h}}_{n,m}^{LS} = \frac{{{{\bf{Y}}_{n,m}}{\bf{s}}_{n,m}^u}}{{\sqrt {{P_t}} U}} = {{\bf{h}}_{n,m}} + {\bf{z}}_{n,m}^{'},
\end{equation}
where ${\bf{z}}_{n,m}^{'} = \frac{{{{\bf{Z}}_{n,m}}{\bf{s}}_{n,m}^u}}{{\sqrt {{P_t}} U}}$ is the transformed Gaussian noise matrix with each element following 
i.i.d. $\mathcal{CN}(0,\frac{{\sigma _n^2}}{{{P_t}}})$. Stack the CSI estimates at ${N_c}$ subcarriers of ${M_s}$ packets to generate ${\bf{\hat H}}_m^{\rm LS} \in {\mathbb{C}^{P \times {N_c}}}$, where ${[ {{\bf{\hat H}}_m^{\rm LS}} ]_{:,n}} = {\bf{\hat h}}_{n,m}^{\rm LS}$. The corresponding actual value to ${\bf{\hat H}}_m^{\rm LS}$ is ${{\bf{H}}_m} \in {\mathbb{C}^{P \times {N_c}}}$, where ${[ {{{\bf{H}}_m}} ]_{:,n}} = {{\bf{h}}_{n,m}}$.

\subsection{FFT-based ISAC Transform} \label{sec:FFT_ISAC}

Let ${{\bf{F}}_N}$ denote a $N$-point discrete Fourier transform (DFT) matrix, we have ${\left[ {{{\bf{F}}_N}} \right]_{{n_1},{n_2}}} = {e^{ - j\frac{{2\pi }}{N}{n_1}{n_2}}}$, where ${n_1},{n_2} = 0,1, \cdots ,N - 1$. Then, the $N$-point inverse discrete Fourier transform (IDFT) matrix is ${\bf{F}}_N^H$.
The FFT-based ISAC transform of ${\bf{\hat H}}_m^{\rm LS}$ can be expressed as~\cite{Sturm2011Waveform}
\begin{equation}\label{equ:FFT_ISAC}
	{\bf{\tilde H}}_m^{\rm LS} = {\bf{T}}\left( {{\bf{\hat H}}_m^{\rm LS}} \right) = {\left[ {{\bf{F}}_{{N_c}}^H{{\left( {{{\bf{F}}_P}{\bf{\hat H}}_m^{\rm LS}} \right)}^T}} \right]^T},
\end{equation}
where ${{\bf{F}}_P}$ and ${{\bf{F}}_{{N_c}}}$ are $P$-point and $N_c$-point DFT matrices, respectively. 

The aforementioned CSI estimates are expressed in the antenna-frequency domain. Based on \eqref{equ:channel_vec}, we have
\begin{equation}\label{equ:CSI_real}
	{\left[ {{{\bf{H}}_m}} \right]_{p,n}}{\rm{ = }}\sum\limits_{l = 0}^{L - 1} {\left[ {{\alpha _{m,l}}{e^{ - j2\pi n\Delta f{\tau _l}}}{e^{j\frac{{2\pi }}{\lambda }(p - 1){d_a}\sin {\theta _{_l}}}}} \right]},
\end{equation}
where ${\alpha _{m,l}} = {b_{C,l}}{e^{j2\pi m{T_s^p}{f_{d,l}}}}$. The $k$th element of the DFT of the $n$th column of ${\bf{\hat H}}_m^{\rm LS}$ is 
\begin{equation}\label{equ:FFT_H}
	\begin{array}{l}
		\begin{array}{l}
			{\left[ {{\bf{\bar h}}_{n,m}^{\rm LS}} \right]_k} = {\left[ {{{\bf{F}}_P}{\bf{\hat h}}_{n,m}^{\rm LS}} \right]_k}\\
			= \sum\limits_{l = 0}^{L - 1} {\sum\limits_{p = 1}^P {{\alpha _{n,m,l}}{e^{j\frac{{2\pi }}{\lambda }\left( {p - 1} \right){d_a}\sin {\theta _l}}}{e^{ - j\frac{{2\pi }}{P}\left( {p - 1} \right)k}}} }  + {z_k},
		\end{array}
	\end{array}
\end{equation}
where ${\alpha _{n,m,l}} = {\alpha _{m,l}}{e^{ - j2\pi n\Delta f{\tau _l}}}$, and ${z_k}$ is the transformed Gaussian noise. It is easily obtained that the modulus of ${\left[ {{\bf{\bar h}}_{n,m}^{\rm LS}} \right]_k}$ is largest when $k_l  = {\frac{{Pd_a\sin {\theta _l}}}{\lambda }}$ in the high SNR regime. Since the index can only be integer, $k_l^\theta  = \left\lfloor {\frac{{Pd_a\sin {\theta _l}}}{\lambda }} \right\rfloor$ should be the maximal point, where $\left\lfloor  \cdot  \right\rfloor$ is rounding-off operator. Similarly, by applying IDFT to each row of ${\bf{\hat H}}_m^{\rm LS}$, the maximal point of each row of the IDFT of ${\bf{\hat H}}_m^{\rm LS}$ is $k_l^\tau  = \left\lfloor {N_c\Delta f{\tau _l}} \right\rfloor $. Therefore, the indices of the FFT-based ISAC transform are changed into the angle-delay domain. Since FFT and IFFT are linear transforms, the inversion of the FFT-based ISAC transform is
\begin{equation}\label{equ:FFT_ISAC_inv}
	{\bf{\hat H}}_m^{\rm LS} = {{\bf{T}}^{ - 1}}\left( {{\bf{\tilde H}}_m^{\rm LS}} \right) = {\bf{F}}_P^H{\left[ {{{\bf{F}}_{{N_c}}}{{\left( {{\bf{\tilde H}}_m^{\rm LS}} \right)}^T}} \right]^T}.
\end{equation}

\begin{figure}[!t]
	\centering
	\includegraphics[width=0.35\textheight]{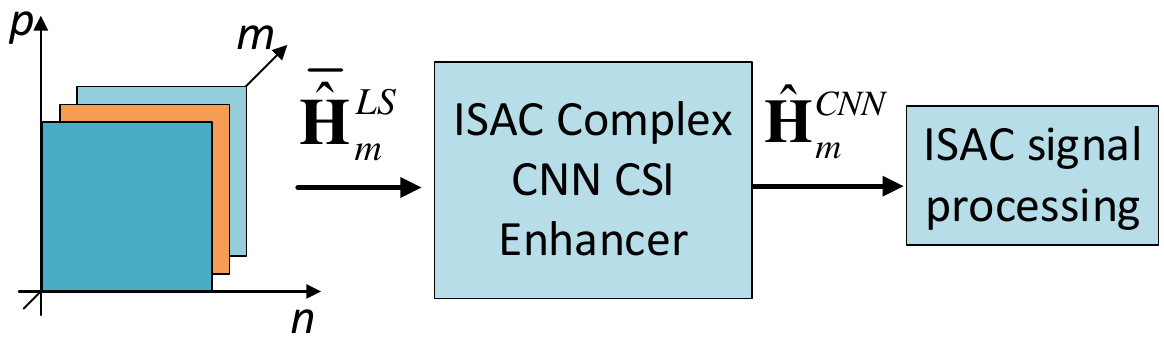}%
	\DeclareGraphicsExtensions.
	\caption{The signal processing diagram of the ISAC system.}
	\label{fig:ISAC diagram}
\end{figure}

Since the channel expression is usually more sparse in the transformed domain, i.e., the angle-delay domain, the FFT-based ISAC transform is useful for feature extraction in the ISAC complex CNN CSI enhancer.

\section{ISAC Complex CNN CSI Enhancer}\label{sec:ISAC_CNN_processing}

The signal processing diagram of the entire ISAC system is shown in Fig.~\ref{fig:ISAC diagram}. The CSI estimate obtained by the LS estimator is first refined by the ISAC complex CNN CSI enhancer. Then, the refined CSI estimates are used to simultaneously demodulate communication signals and estimate sensing parameters. Since the sensing parameters are contained in the CSI, if we can reduce the MSE of the refined CSI to an extremely low level, the estimation accuracy of AoA and range should be improved. Therefore, our optimization objective for the ISAC complex CNN CSI enhancer is to minimize the MSE of the refined CSI, which can unify the communication and sensing objectives.

In this section, we focus on the design of the ISAC complex CNN CSI enhancer, and the communication demodulation and sensing processing will be presented in the next section. We first introduce the CNN structure and training processes in detail and then analyze the complexity of the proposed CSI enhancer.

\subsection{Complex-valued Computation Layers for CNN}
Three kinds of layers are used in the ISAC complex CNN: complex linear, complex activation, and complex convolutional layers. The prominent feature of the complex-valued neural network is that the weights and inputs are both complex values, and the key to achieving the complex neural network is to apply the complex multiplication operation to all the network layers.

\subsubsection{Complex Linear Layer}

The weights and bias parameters of the complex linear layer are ${\bf{W}} = {{\bf{W}}_r} + j{{\bf{W}}_i}$ and ${\bf{b}} = {{\bf{b}}_r} + j{{\bf{b}}_i}$, respectively. With the input formed as ${\bf{x}} = {{\bf{x}}_r} + j{{\bf{x}}_i}$, we can express the output of the complex linear layer as:
\begin{equation}\label{equ:complex_linear}
	{{\bf{y}}_L} = L\left( {{\bf{x}};{\bf{W}},{\bf{b}}} \right) = {\bf{Wx}} + {\bf{b}}.
\end{equation}
Based on the complex multiplication operation, we obtain
\begin{equation}\label{equ:complex_linear_ex}
	{{\bf{y}}_L} = {{\bf{W}}_r}{{\bf{x}}_r} - {{\bf{W}}_i}{{\bf{x}}_i} + {{\bf{b}}_r} + j\left( {{{\bf{W}}_r}{{\bf{x}}_i} + {{\bf{W}}_i}{{\bf{x}}_r} + {{\bf{b}}_i}} \right),
\end{equation}

We can see that the complex layer output can be deterministically calculated by real-value linear layers, which can be expressed as 
\begin{equation}\label{equ:complex_linear_eq}
	\begin{aligned}
		{{\bf{y}}_L} = &L\left( {{{\bf{x}}_r};{{\bf{W}}_r},{\bf{0}}} \right) + L\left( {{{\bf{x}}_i}; - {{\bf{W}}_i},{\bf{0}}} \right) + {{\bf{b}}_r}\\
		&+ j\left[ {L\left( {{{\bf{x}}_i};{{\bf{W}}_r},{\bf{0}}} \right) + L\left( {{{\bf{x}}_r}; {{\bf{W}}_i},{\bf{0}}} \right) + {{\bf{b}}_i}} \right].
	\end{aligned}
\end{equation}
Based on \eqref{equ:complex_linear_eq}, we can construct the complex linear layer operation with two real-valued linear layers. 

\subsubsection{Complex Convolutional Layer}

\begin{figure}[!t]
	\centering
	\includegraphics[width=0.30\textheight]{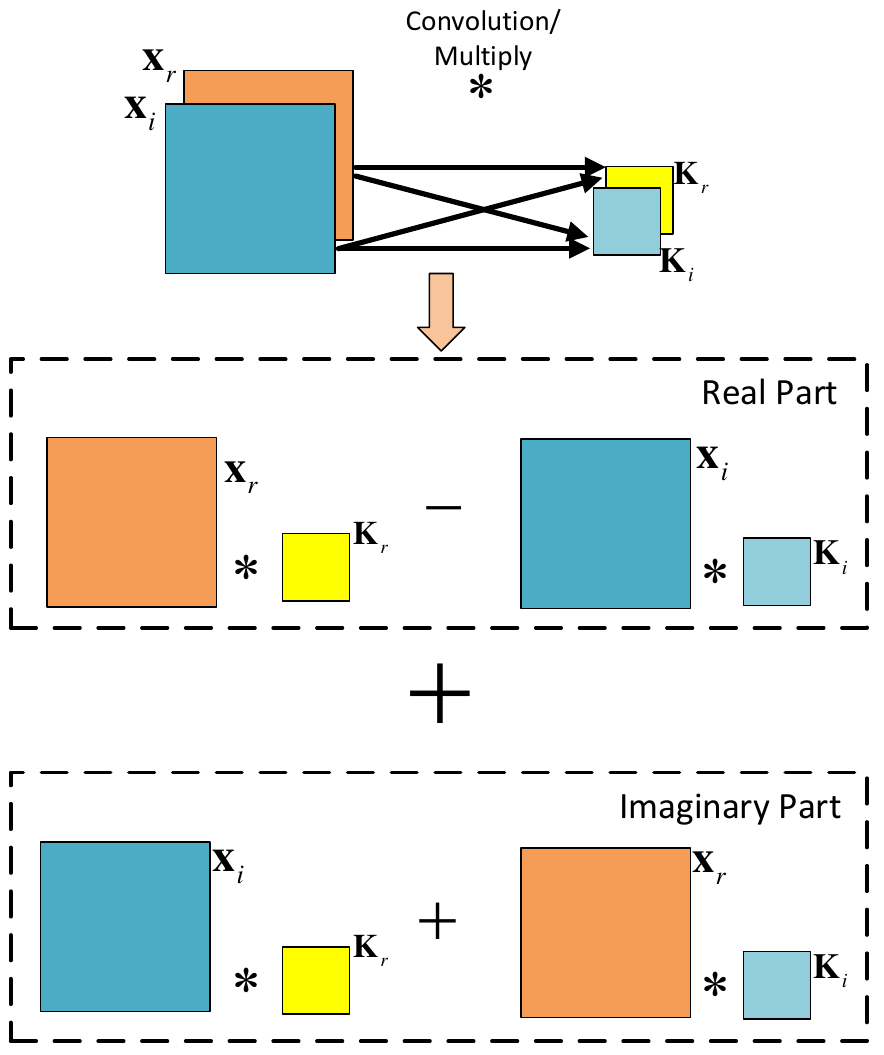}%
	\DeclareGraphicsExtensions.
	\caption{The illustration of complex-valued linear and convolution layers.}
	\label{fig:Complex_Linear}
\end{figure}

Let ${\bf{K}} = {{\bf{K}}_r} + j{{\bf{K}}_i}$ denote the convolution kernel. Since the convolution operation is also a linear transform, the output of complex-valued convolution can be represented by
\begin{equation}\label{equ:complex_conv}
	\begin{aligned}
		{f_C}\left( {{\bf{x}};{\bf{K}}} \right) &= {\bf{K}} * {\bf{x}}\\
		&= {{\bf{K}}_r} * {{\bf{x}}_r} - {{\bf{K}}_i} * {{\bf{x}}_i} + j\left( {{{\bf{K}}_r} * {{\bf{x}}_i} + {{\bf{K}}_i} * {{\bf{x}}_r}} \right),
	\end{aligned}
\end{equation}
where $*$ is the convolutional operator. Based on \eqref{equ:complex_conv}, we can use two real-valued convolutional layers to deterministically obtain the complex-valued convolution results. 

The aforementioned complex-valued linear computation layers are shown in Fig.~\ref{fig:Complex_Linear}, which is similar to the butterfly computation structure.

\subsubsection{Complex Activation Layer}
The activation function is used to add non-linearity into the forward propagation. According to~\cite{trabelsi2017deep}, we apply the leaky Relu function to both real and imaginary parts of input to construct the activation layer, which can be expressed as
\begin{equation}\label{equ:complex_activation}
	{\rm{CLRelu}}\left( {\bf{y}} \right){\rm{ = LeakyRelu}}\left( {{{\bf{y}}_r}} \right){\rm{ + }}j{\rm{LeakyRelu}}\left( {{{\bf{y}}_i}} \right),
\end{equation}
where ${{{\bf{y}}_r}}$ and ${{{\bf{y}}_i}}$ are the real and imaginary parts of ${\bf{y}}$, respectively, and ${\rm{LeakyRelu}}\left( y \right) = \left\{ \begin{array}{l}
	y,y \ge 0,\\
	ay,y < 0,
\end{array} \right.$ is the real-valued leaky Relu function with $a$ being a small positive value.

\begin{figure}[!t]
	\centering
	\includegraphics[width=0.30\textheight]{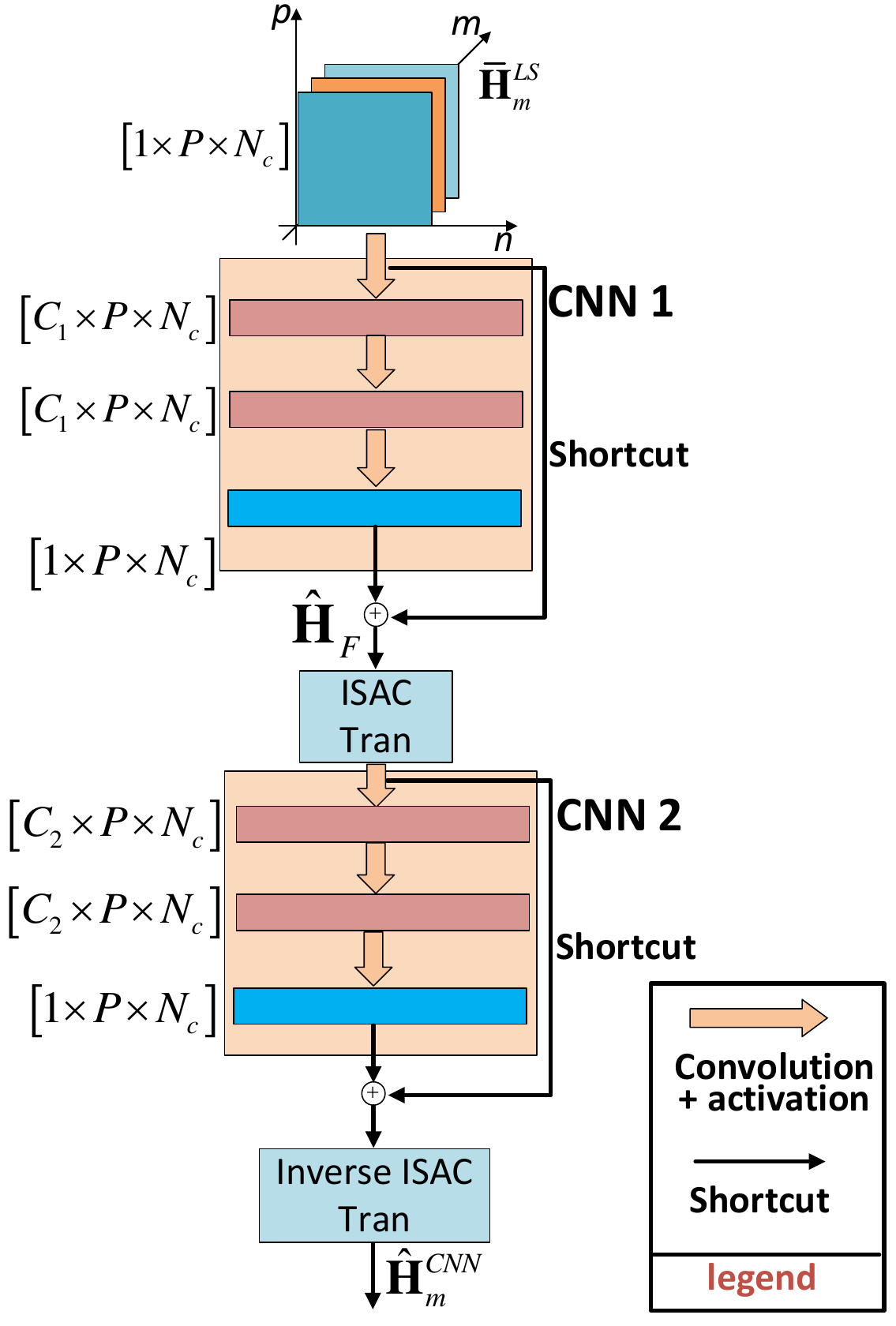}%
	\DeclareGraphicsExtensions.
	\caption{The diagram of ISAC complex CNN.}
	\label{fig:CNN_diagram}
\end{figure}

\subsection{Structure of ISAC Complex CNN}

The diagram of the ISAC complex CNN is shown in Fig.~\ref{fig:CNN_diagram}. We use two Resnet-like CNN structures to construct the backbone of the complex CNN. The general formulation of the output of a complex CNN is
\begin{equation}\label{equ:fCNN}
	{{\bf{\hat H}}_{out}} = {f_{\rm CNN}}\left( {{{{\bf{\hat H}}}_{in}};\Theta } \right),
\end{equation}
where $\Theta $ is the set of all the weights and bias parameters of the complex CNN; ${{\bf{\hat H}}_{in}}$ and ${{\bf{\hat H}}_{out}}$ are the input and output tensors, respectively. In this paper, we use three convolutional layers for each CNN block, and each convolutional layer is the concatenation of complex convolutional and leaky Relu functions, i.e., 
\begin{equation}\label{equ:CNN_1}
	f_{\rm CNN}^1\left( {{{{\bf{\hat H}}}_{in}};{\Theta^1}} \right) = {\rm{CLRelu}}\left( {{f_{\rm CNN}}\left( {{{{\bf{\hat H}}}_{in}};{\Theta^1}} \right)} \right),
\end{equation}
where ${\Theta^1}$ is the parameter set. Let ${f_{\rm CNN}^{b}}(\cdot)$ denote the comprehensive function of a CNN block, which is the concatenation of three $f_{\rm CNN}^1(\cdot)$.
Moreover, the shortcut path is a single-layer complex convolutional layer, and its output is expressed as
\begin{equation}\label{equ:Res_output}
	{{\bf{H}}_{{\rm{Res}}}} = f_{\rm CNN}\left( {{{{\bf{\hat H}}}_{in}};{\Theta _{{\rm{Res}}}}} \right),
\end{equation}
where ${\Theta _{{\rm{Res}}}}$ is the parameter set of the shortcut path.
CNN 1 is used to pre-process the input by extracting the superimposed main features of the initial CSI estimates, and the output of CNN 1 is expressed as
\begin{equation}\label{equ:CNN1_output}
	{{\bf{\hat H}}_F} = {f_{\rm CNN}^{b}}\left( {{\bf{\bar H}}_m^{\rm LS};{\Theta _1}} \right) + f_{\rm CNN}\left( {{\bf{\bar H}}_m^{\rm LS};\Theta _1^{'}} \right),
\end{equation}
where ${\Theta _1}$ and $\Theta _1^{'}$ are the weights parameters of CNN 1 and its shortcut, respectively. Moreover, ${\bf{\bar H}}_m^{\rm LS}$ is the normalization of the initial CSI estimate, ${\bf{\hat H}}_m^{\rm LS}$, and the normalization method will be presented in the next subsection.

The ISAC transform module transforms the feature map output of CNN 1 into the sparse angle-delay domain so that CNN 2 can learn the sparse feature of CSIs on the transformed domain. Finally, the inverse ISAC transform module restores the CSI estimates to their original domain. The output of CNN 2 after the inverse ISAC transform is expressed as
\begin{equation}\label{equ:CNN2_output}
	\begin{array}{l}
		{\bf{\hat H}}_m^{\rm CNN}\\
		= {{\bf{T}}^{ - 1}}\left[ {{f_{\rm CNN}^{b}} \left( {{\bf{T}}\left( {{{{\bf{\hat H}}}_F}} \right);{\Theta _2}} \right) + f_{\rm CNN}\left( {{\bf{T}}\left( {{{{\bf{\hat H}}}_F}} \right);\Theta _2^{'}} \right)} \right],
	\end{array}
\end{equation}
where ${\Theta _2}$ and $\Theta _2^{'}$ are the parameters of CNN 2 and its shortcut, respectively, and ${\bf{T}}\left(  \cdot  \right)$ and ${{\bf{T}}^{ - 1}}\left(  \cdot  \right)$ are FFT-based ISAC transform and inverse transform, respectively.

The dimension of a CSI tensor is denoted by $\left[ {C \times P \times {N_c}} \right]$, where $C$ is the channel number of the input tensor, $P$ is the number of antennas, and $N_c$ is the number of subcarriers. The output of each complex CNN has the same channel number as the input, while the intermediate tensor output can have different channel numbers. The channel numbers of the intermediate output of CNNs 1 and 2 are $C_1$ and $C_2$, respectively. In this paper, the input and output CSI tensors are all with 1 channel. The dimension of the convolution kernel is $3 \times 3$ for both CNNs 1 and 2.

\subsection{Training Methods of ISAC complex CNN CSI Enhancer}
We further introduce the training method of the ISAC complex CNN, including the input normalization and loss function for backpropagation (BP).

\subsubsection{The normalization of input} \label{sec:normalize_input}
Since the raw CSI data usually has a small amplitude, we need to normalize the initial CSI estimation, ${\bf{\hat H}}_m^{\rm LS}$, to avoid the training stagnation due to the gradient vanishment in the training process. We aim to normalize the power of the useful signal, i.e., the true CSI, ${{\bf{H}}_m}$. 

We can exploit the eigenvalues of the autocorrelation of ${\bf{\hat H}}_m^{\rm LS}$ to estimate the power of the useful signal and the noise variance. The autocorrelation of ${\bf{\hat H}}_m^{\rm LS}$ is denoted by ${\bf{R}}_m^{\rm LS} = \frac{1}{{{N_c}}}{\bf{\hat H}}_m^{\rm LS}{\left( {{\bf{\hat H}}_m^{\rm LS}} \right)^H}$. Let ${{\bf{v}}_\Sigma }$ denote the vector composed of the eigenvalues of ${\bf{R}}_m^{\rm LS}$ with the descending order. According to the feature of eigenvalue decomposition~\cite{2023XuJCAS}, we have
\begin{equation}\label{equ:eigenvalue}
	{\left[ {{{\bf{v}}_\Sigma }} \right]_i} = \left\{ \begin{array}{l}
		{\rho _i} + \sigma _N^2,i \le L,\\
		\sigma _N^2,i > L,
	\end{array} \right.
\end{equation}
where $L$ is the number of strong paths, ${\rho _i}$ is the power gain of the $i$th path, and $\sigma _N^2 = \frac{{\sigma _n^2}}{{{P_t}}}$ is the variance of CSI estimation error. The estimation of $L$, denoted by $\hat L$, can be obtained following the procedures in \textbf{Appendix~\ref{appendix:L_est}}. After gaining $\hat L$, the estimation of $\sigma _N^2$ can be expressed as
\begin{equation}\label{equ:Noise_H}
	\hat \sigma _N^2 = \frac{{\sum\limits_{i = \hat L + 1}^P {{{\left[ {{{\bf{v}}_\Sigma }} \right]}_i}} }}{{P - \hat L}}.
\end{equation}
Then, the power of useful signal in ${\bf{\hat H}}_m^{\rm LS}$ can be estimated as
\begin{equation}\label{equ:pwoer_gain}
	\rho _h^2 = \sum\limits_{i = 1}^{i = \hat L} {\left( {{{\left[ {{{\bf{v}}_\Sigma }} \right]}_i} - \hat \sigma _N^2} \right)}.
\end{equation}
Finally, the normalization of ${\bf{\hat H}}_m^{\rm LS}$ is given by
\begin{equation}\label{equ:normalize_H}
	{\bf{\bar H}}_m^{\rm LS} = \frac{{{\bf{\hat H}}_m^{\rm LS}}}{{\sqrt {\rho _h^2} }} = \frac{{{{\bf{H}}_m}}}{{\sqrt {\rho _h^2} }}{\rm{ + }}{\bf{Z}}_m^{'},
\end{equation}
where ${\bf{Z}}_m^{'}$ is the noise term with ${\left[ {{\bf{Z}}_m^{'}} \right]_{:,n}} = \frac{{{\bf{z}}_{n,m}^{'}}}{{\sqrt {\rho _h^2} }}$. Note that due to the normalization procedure, when generating the training data, the true values of CSI should be generated as $\frac{{{{\bf{H}}_m}}}{{\sqrt {\rho _h^2} }}$.

\subsubsection{Loss Function}
Since the input and output of CNN are both complex values, we define the mean square error loss function for complex values as
\begin{equation}\label{equ:Loss}
	\begin{array}{l}
		J\left( {{{\bf{H}}_m},{\bf{\hat H}}_m^{\rm CNN}} \right) = \left\| {{{\bf{H}}_m} - {\bf{\hat H}}_m^{\rm CNN}} \right\|_F^2\\
		= {\rm{Tr}}\left\{ {{{\left( {{{\bf{H}}_m} - {\bf{\hat H}}_m^{\rm CNN}} \right)}^H}\left( {{{\bf{H}}_m} - {\bf{\hat H}}_m^{\rm CNN}} \right)} \right\},
	\end{array}
\end{equation}
where ${\rm{Tr}}\left(  \cdot  \right)$ is the operation to derive the trace of a matrix. Using BP to find the optimal parameters by minimizing $J\left( {{{\bf{H}}_m},{\bf{\hat H}}_m^{\rm CNN}} \right)$, we can finally obtain the optimal parameters of ISAC complex CNN CSI enhancer, which is expressed as
\begin{equation}\label{equ:minimize_J}
	\begin{array}{l}
		{\hat \Theta _1},{\hat \Theta _2},\hat \Theta _1^{'},\hat \Theta _2^{'} = \mathop {\arg \min }\limits_{{\Theta _1},{\Theta _2},\Theta _1^{'},\Theta _2^{'}} J\left( {{{\bf{H}}_m},{\bf{\hat H}}_m^{\rm CNN}} \right).
	\end{array}
\end{equation}

\subsection{Complexity Analysis and Comparison}
In this subsection, we analyze the complexity of the above ISAC complex CNN CSI enhancer and compare it with the LS and linear minimum mean square error (LMMSE) CSI estimation methods. 

\subsubsection{Complexity of the ISAC complex CNN CSI enhancer}
The computation complexity of the ISAC complex CNN CSI enhancer mainly comes from two CNNs, two FFT-based ISAC transforms, and the procedure of computing the eigenvalues of ${\bf{R}}_m^{\rm LS}$ in Section~\ref{sec:normalize_input}. The aggregate complexity of two CNNs is $\mathcal{O}\left\{ {{3^2} \times P{N_c} \times \left[ {2\left( {{C_1} + {C_2}} \right) + \left( {{C_1}^2 + {C_2}^2} \right)} \right]} \right\}$, the aggregate complexity of two FFT-based transforms is $\mathcal{O}\left\{ {2 \times P{N_c}\log \left( {P{N_c}} \right)} \right\}$, and the complexity of computing the eigenvalues of ${\bf{R}}_m^{\rm LS}$ is $\mathcal{O}\{P^3\}$. Therefore, the comprehensive complexity of the ISAC complex CNN CSI enhancer is $\mathcal{O}\left\{ {P{N_c}\left[ {9\left( {{C_1}^2 + {C_2}^2} \right) + 2{{\log }_2}\left( {P{N_c}} \right)} \right]} + P^3 \right\}$. 

\subsubsection{Complexity of the LS method}
The complexity of the LS method comes from the complex-valued division, which is $\mathcal{O}(PN_c)$.

\subsubsection{Complexity of the LMMSE method}
The CSI estimation of the LMMSE method at the $n$th subcarrier of the $m$th packet can be expressed as~\cite{2010MIMO}
\begin{equation}\label{equ:MMSE}
	{\bf{\hat h}}_{n,m}^{\rm MMSE} = {{\bf{R}}_{{\bf{hh}}}}{\left[ {{{\bf{R}}_{{\bf{hh}}}} + \hat \sigma _N^2{\bf{I}}} \right]^{ - 1}}{\bf{\hat h}}_{n,m}^{LS},
\end{equation}
where ${{\bf{R}}_{{\bf{hh}}}} = E\left( {{{\bf{h}}_{n,m}}{\bf{h}}_{n,m}^H} \right)$ is the expectation of the autocorrelation of ${\bf{h}}_{n,m}$. 

The LMMSE method adds the matrix inversion and multiplication operations based on the LS method. Therefore, the complexity of the LMMSE method is $\mathcal{O}[2(P^2 + P^3) N_c]$. 

Since $C_1$ and $C_2$ are usually smaller than 3, the complexity of the ISAC complex CNN CSI enhancer is between those for the LS and LMMSE methods.

\begin{figure}[!t]
	\centering
	\includegraphics[width=0.35\textheight]{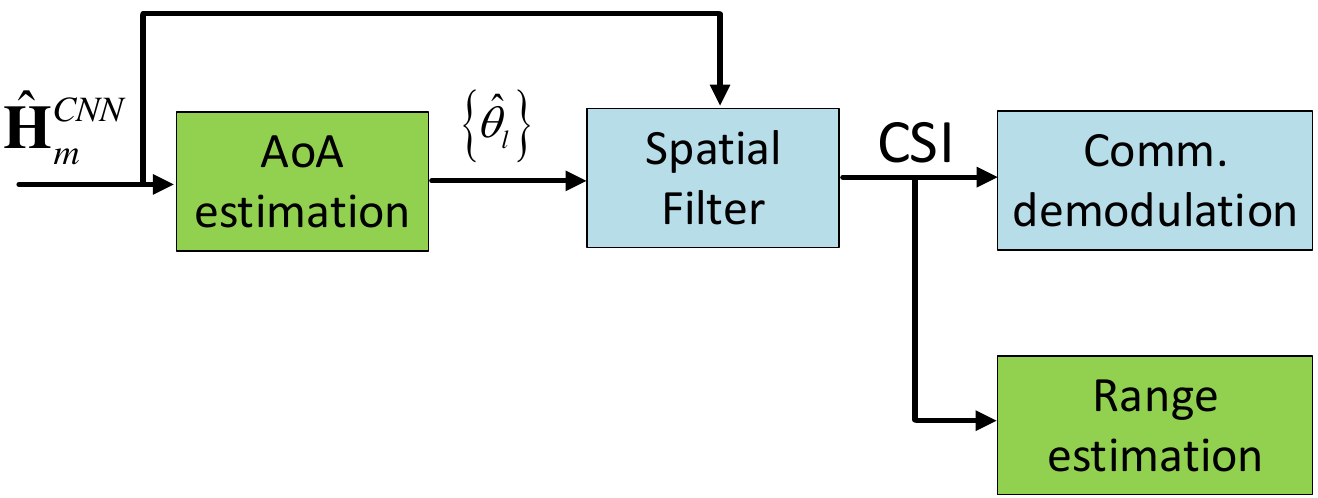}%
	\DeclareGraphicsExtensions.
	\caption{The ISAC signal processing diagram.}
	\label{fig:ISAC sigprocessing}
\end{figure}

\section{ISAC Signal Processing}\label{sec:ISAC_processing}
In this section, we introduce the ISAC signal processing methods. We first estimate the AoA based on the CSI estimation enhanced by the ISAC complex CNN, ${\bf{\hat H}}_m^{\rm CNN}$, and then apply the estimated AoA as the prior information to generate a baseband spatial filter (SF) based on beamforming (BF). The SF is applied to the enhanced CSI estimates and the filtered CSI is used for communication demodulation and estimating the ranges of paths with different AoAs. Finally, we propose a novel biased FFT-based sensing method for range sensing.

\subsection{AoA Estimation and Spatial Filtering}

The enhanced CSI estimation, ${\bf{\hat H}}_m^{\rm CNN}$, can be expressed as
\begin{equation}\label{equ:H_CNN}
	{\bf{\hat H}}_m^{\rm CNN} = \frac{{{{\bf{H}}_m}}}{{\sqrt {\rho _h^2} }} {\rm{ + }}{\bf{\tilde Z}}_m^{'},
\end{equation}
where ${\bf{\tilde Z}}_m^{'}$ is the noise matrix suppressed by the ISAC CNN CSI enhancer, and ${{\bf{H}}_m}$ is the true CSI that contains the AoA and range of UE and scatterers. We first use the multiple signal classification (MUSIC) method to estimate AoA based on ${\bf{\hat H}}_m^{\rm CNN}$. 

\subsubsection{AoA Estimation} \label{sec:AoA_est}
By applying eigenvalue decomposition to the autocorrelation of ${\bf{\hat H}}_m^{\rm CNN}$, we obtain 
\begin{equation}\label{equ:eig_MUSIC}
	\left[ {{{\bf{U}}_a},{{\bf{\Sigma }}_a}} \right] = {\rm{eig}}\left\{ {\frac{1}{{{N_c}}}{\bf{\hat H}}_m^{\rm CNN}{{\left( {{\bf{\hat H}}_m^{\rm CNN}} \right)}^H}} \right\},
\end{equation}
where ${\rm{eig}}(\cdot)$ denotes eigenvalue decomposition, ${{\bf{\Sigma }}_a}$ is the eigenvalue matrix with diagonal elements being the eigenvalues in descending order, and ${{\bf{U}}_a}$ is the eigenmatrix with each column being the eigenvector corresponding to the eigenvalue. Since the number of strong paths has been estimated as $\hat L$ in subsection~\ref{sec:normalize_input}. Therefore, the noise subspace can be obtained as ${{\bf{U}}_N} = {\left[ {{{\bf{U}}_a}} \right]_{:,\hat L + 1:P}}$. According to \cite{2023XuJCAS}, the angle spectrum function is given by
\begin{equation}\label{equ:f_a}
	{f_a}\left( {\theta ;{{\bf{U}}_N}} \right) = {{\bf{a}}^H}\left( \theta  \right){{\bf{U}}_N}{\left( {{{\bf{U}}_N}} \right)^H}{\bf{a}}(\theta ).
\end{equation}
The minimal points of ${f_a}\left( {\theta ;{{\bf{U}}_N}} \right)$ are the AoA estimates. Using the two-step Newton descending method in \cite{2023XuJCAS}, we can obtain the minimal points of ${f_a}\left( {\theta ;{{\bf{U}}_N}} \right)$. 

The set of AoA estimation is denoted by ${\Theta _A} = {\left. {\{ {{{\hat \theta }_l}} \}} \right|_{l = 0,1,...,\hat L - 1}}$, and the points in ${\Theta _A}$ are sorted by the ascending order of ${f_a}\left( {\theta ;{{\bf{U}}_N}} \right)$. 

\subsubsection{Spatial Filtering} \label{sec:SF}

By stacking the steering vectors of all the estimated AoAs in ${\Theta _A}$, we obtain the estimated steering matrix as
\begin{equation}\label{equ:A_theta}
	{\bf{A}}\left( {{\Theta _A}} \right) = \left[ {{\bf{a}}({\theta _0}),{\bf{a}}({\theta _1}),...,{\bf{a}}({\theta _{\hat L - 1}})} \right],
\end{equation}
where ${\bf{a}}({\theta _l})$ is given in \eqref{equ:phase_difference}. Using the low-complexity LS beamforming method, we can obtain the receiving BF vector for the signals with AoA ${\hat \theta _l}$ as 
\begin{equation}\label{equ:w_Rl}
	{{\bf{w}}_{R,l}} = {\left[ {{{\bf{A}}^H}\left( {{\Theta _A}} \right)} \right]^\dag }{{\bf{i}}_l} \in {\mathbb{C}^{P \times 1}},
\end{equation}
where ${\left[  \cdot  \right]^\dag }$ is the pseudo-inverse operation to a matrix, and ${{\bf{i}}_l}$ is the one-hot vector with only the $l$th element being 1. Using ${{\bf{w}}_{R,l}}$ to spatially filter ${\bf{\hat H}}_m^{\rm CNN}$, we obtain the filtered CSI as 
\begin{equation}\label{equ:h_Rl}
	{{\bf{\hat h}}_{R,l}} = {\left( {{{\bf{w}}_{R,l}}} \right)^H}{\bf{\hat H}}_m^{\rm CNN} \in {\mathbb{C}^{1 \times {N_c}}}.
\end{equation}

Combining \eqref{equ:H_CNN} and \eqref{equ:h_Rl}, we can obtain the $n$th element of ${{\bf{\hat h}}_{R,l}}$ can be expressed as
\begin{equation}\label{equ:h_Rl_n}
	{\left[ {{{{\bf{\hat h}}}_{R,l}}} \right]_n} = {\alpha _{n,m,l}}{e^{ - j2\pi n\Delta f{\tau _l}}} + {\tilde z_{n,m,l}},
\end{equation}
where ${\alpha _{n,m,l}} = {b_{C,l}}{e^{j2\pi m{T_s^p}{f_{d,l}}}}{\bf{w}}_{R,l}^H{\bf{a}}\left( {{\theta _l}} \right)$ is the useful power gain, and ${\tilde z_{n,m,l}}$ is the term of noise and interference. Since ${{\bf{w}}_{R,l}}$ is the directional receiving BF vector, $E\left( {\left\| {{{\tilde z}_{n,m,l}}} \right\|_2^2} \right)$ should be much smaller than $E\left( {\left\| {{\alpha _{n,m,l}}} \right\|_2^2} \right)$ in the high SNR regime.

\subsection{Biased FFT-based Range Sensing}

Based on \eqref{equ:h_Rl_n}, we can see that ${{\bf{\hat h}}_{R,l}}$ contains the delay of the $l$th path, and we can thereby estimate the range based on ${{\bf{\hat h}}_{R,l}}$. As aforementioned, the FFT-based ISAC transform in Section \ref{sec:FFT_ISAC} can be used to transform ${{\bf{\hat h}}_{R,l}}$ into the delay domain. However, it can only give the discrete spectrum with interval $\Delta r = \frac{c}{{{N_c}\Delta f}}$, and the range estimation accuracy is thus restricted by $\Delta r$, especially when the true range is at the midpoint of an interval. 

To resolve this problem, we propose a novel biased FFT-based sensing method by actively introducing multiple phase biases to the CSI and averaging the debiased sensing results estimated by a simple FFT-based sensing method to obtain a more accurate estimation of the sensing parameter.

Then, we present the derivation of the biased FFT-based sensing method. We use ${r_t}$ to denote the true value of the range, and we have ${r_1} < {r_t} < {r_2}$ and ${r_2} = {r_1} + \Delta r$. The discrete spectrum obtained by applying FFT to the raw sequence is the uniform sampling of the discrete-time Fourier transform (DTFT) spectrum, and the amplitude spectrum of DTFT should be symmetric with respect to the line $r = {r_t}$ in the high SNR regime~\cite{Zhang2023}. Therefore, the range estimate of the FFT-based sensing method should be the grid point that is closer to $r_t$, which can be expressed as
\begin{equation}\label{equ:r_est}
	{r_{est}} = \mathop {\arg \min }\limits_{r \in \left\{ {{r_1},{r_2}} \right\}} \left| {{r_t} - r} \right|,
\end{equation}
and the error should be
\begin{equation}\label{equ:r_error}
	r_e = \mathop {\min }\limits_{r \in \left\{ {{r_1},{r_2}} \right\}} \left| {{r_t} - r} \right|.
\end{equation}

According to \eqref{equ:h_Rl_n}, by multiplying ${e^{-j2\pi n\Delta f\frac{{k {r_\delta }}}{c}}}$ to the $n$th element of ${{\bf{\hat h}}_{R,l}}$, we obtain ${{\bf{\hat h}}_{R,l,k}} = {{\bf{\hat h}}_{R,l}} \odot {{\bf{h}}_k}$, where ${\left[ {{{\bf{h}}_k}} \right]_n} = {e^{-j2\pi n\Delta f\frac{{k{r_\delta }}}{c}}}$, and $ \odot $ is the Hadamard product operator. The $n$th element of ${{\bf{\hat h}}_{R,l,k}}$ is expressed as
\begin{equation}\label{equ:r_h_Rlk}
	{\left[ {{{{\bf{\hat h}}}_{R,l,k}}} \right]_n} = {\alpha _{n,m,l}}{e^{ - j2\pi n\Delta f\left( {{\tau _l} + \frac{{k{r_\delta }}}{c}} \right)}} + {\tilde z_{n,m,l}}{e^{-j2\pi n\Delta f\frac{{k{r_\delta }}}{c}}}.
\end{equation}
We can see that the true value of the range contained in ${{\bf{\hat h}}_{R,l,k}}$ is biased by $k{r_\delta }$. Therefore, the range estimates based on ${{\bf{\hat h}}_{R,l,k}}$ should be debiased by $k{r_\delta }$, i.e., ${\bar r_{est,k}} = {r_{est}} - k{r_\delta }$. 

Split the interval $\left[ {{r_t} - \frac{{\Delta r}}{2}, {r_t}{\rm{ + }}\frac{{\Delta r}}{2}} \right]$ into ${N_r}$ even pieces with grid spacing ${r_\delta } = \frac{{\Delta r}}{{{N_r}}}$. Using \eqref{equ:r_h_Rlk} to add biases to the true value $r_t$, we obtain a set of biased ranges, denoted by $\{{r_k}\} = {\left. {\{ {{r_t} + k{r_\delta }} \}} \right|_{k \in [ - \frac{{{N_r}}}{2},\frac{{{N_r}}}{2}]}}$, where $k$ is integer. Assume ${r_{est}} = {r_1}$ for $k \in [ { - \frac{{{N_r}}}{2},\bar k} ]$, and ${r_{est}} = {r_2}$ for $k \in [ {\bar k + 1,\frac{{{N_r}}}{2}} ]$. The average of all the debiased range estimates can be expressed as
\begin{equation}\label{equ:r_est_bias}
	\begin{array}{l}
		{{\hat r}_{est}} = \frac{{\sum\limits_{k =  - {N_r}/2}^{\bar k} {\left( {{r_1} - k{r_\delta }} \right)}  + \sum\limits_{k = \bar k + 1}^{{N_r}/2} {\left( {{r_2} - k{r_\delta }} \right)} }}{{{N_r} + 1}}\\
		= \frac{{\sum\limits_{k =  - {N_r}/2}^{\bar k} {\left( {{r_1} - k{r_\delta }} \right)}  + \sum\limits_{k = \bar k + 1}^{{N_r}/2} {\left( {{r_1} + {N_r}{r_\delta } - k{r_\delta }} \right)} }}{{{N_r} + 1}}\\
		= {r_1} + {r_\delta }\left[ {\frac{{{N_r}\left( {{N_r}/2 - \bar k} \right)}}{{{N_r} + 1}}} \right].
	\end{array}
\end{equation}

We prove that the error of biased FFT-based sensing, i.e., $\left| {{{\hat r}_{est}} - {r_t}} \right|$, is always smaller than $\left| {{r_{est}} - {r_t}} \right|$ in \textbf{Appendix~\ref{appendix:proof}}.

The procedure of biased FFT-based sensing is summarized in \textbf{Algorithm~\ref{alg:biasedFFT}}. The biased FFT-based sensing method improves sensing accuracy in a mechanism similar to the diversity gain, as will be demonstrated in Section~\ref{sec:sensing_performance}.

\subsubsection{Complexity of Biased FFT-based Sensing Method}

The complexity of the proposed biased FFT-based sensing method is mainly from the $N_r$ rounds of FFT-based sensing. Therefore, the complexity of the proposed biased FFT-based sensing method is $O\left\{ {N_r {N_c}\log \left( {{N_c}} \right)} \right\}$, which is significantly lower than the subspace-based sensing method, such as the MUSIC method whose typical complexity is about $O\left\{ {{N_c}^3 + {N_c}^2} \right\}$.

\begin{algorithm}[!t]
	\caption{Biased FFT-based Sensing Method}
	\label{alg:biasedFFT}
	\KwIn{Spatially filtered CSI vector: ${{\bf{\hat h}}_{R,l}}$; The grid interval: $\Delta r = \frac{c}{{{N_c}\Delta f}}$; The number of even pieces for each interval: ${N_r}$.
	}
	\KwOut{The range estimation ${\hat r_{est}}$.}
	\textbf{Initialize: } \\
	1) Calculate ${r_\delta } = \frac{{\Delta r}}{{{N_r}}}$.\\
	2) Generate the range bias set ${\left. {\left\{ {k{r_\delta }} \right\}} \right|_{k \in \left[ { - \frac{{{N_r}}}{2},\frac{{{N_r}}}{2}} \right]}}$.\\
	\textbf{Process: } \\
	\textbf{Step 1}: Add bias phase terms to ${{\bf{\hat h}}_{R,l}}$ according to \eqref{equ:r_h_Rlk}, which generates ${N_r} + 1$ biased vectors ${ {\{ {{{{\bf{\hat h}}}_{R,l,k}}} \}} |_{k \in [ { - \frac{{{N_r}}}{2},\frac{{{N_r}}}{2}} ]}}$.
	
	\textbf{Step 2}: Apply FFT-based ISAC range sensing method to each vector in ${ {\{ {{{{\bf{\hat h}}}_{R,l,k}}} \}} |_{k \in [ { - \frac{{{N_r}}}{2},\frac{{{N_r}}}{2}} ]}}$, and generate the corresponding range estimates ${\left. {\left\{ {{{\hat r}_k}} \right\}} \right|_{k \in \left[ { - \frac{{{N_r}}}{2},\frac{{{N_r}}}{2}} \right]}}$.
	
	\textbf{Step 3}: Debias the range estimation and generate ${\left. {\left\{ {{{\hat r}_{est,k}} = {{\hat r}_k} - k{r_\delta }} \right\}} \right|_{k \in \left[ { - \frac{{{N_r}}}{2},\frac{{{N_r}}}{2}} \right]}}$.
	
	\textbf{Step 4}: Calculate the average of the debiased range estimates: ${\hat r_{est}} = \frac{{\sum\limits_{k =  - {N_r}/2}^{{N_r}/2} {{{\hat r}_{est,k}}} }}{{{N_r} + 1}}$.
	
\end{algorithm}

\subsection{Communication Processing}

The received signals of the $n$th subcarrier at the $m$th OFDM symbol can be expressed as
\begin{equation}\label{equ:y_nm_C_vec}
	{\bf{y}}_{n,m}^C = \sqrt {{P_t}} {d_{n,m}}{{\bf{h}}_{n,m}} + {{\bf{z}}_{n,m}},
\end{equation}
where ${d_{n,m}}$ is the transmitted data at the $n$th subcarrier of the $m$th OFDM symbol, and ${{\bf{z}}_{n,m}}$ is the Gaussian noise vector with each element following $\mathcal{CN}(0,{\sigma _n^2})$. 

Here, the CSI estimate obtained by the CNN CSI enhancer corresponding to ${{\bf{h}}_{n,m}}$ is ${\bf{\hat h}}_{n,m}^{\rm CNN} = {\left[ {{\bf{\hat H}}_m^{\rm CNN}} \right]_{:,n}}$.
Using the maximum likelihood criterion, we estimate the communication data as
\begin{equation}\label{equ:d_nm_est}
	{\hat d_{n,m}} = \mathop {\arg \min }\limits_{{d_\Theta } \in {\Theta _{\rm QAM}}} {\left| {\frac{{{{\left( {{\bf{\hat h}}_{n,m}^{\rm CNN}} \right)}^\dag }{\bf{y}}_{n,m}^C}}{{\sqrt {{P_t}} }} - {d_\Theta }} \right|^2},
\end{equation}
where ${\Theta _{\rm QAM}}$ is the used quadrature amplitude modulation (QAM) constellation.

\section{Simulation Results}\label{sec:Simulation}
In this section, we present the normalized MSE (NMSE) of the CSI estimate obtained by the ISAC complex CNN CSI enhancer, the BER performance of demodulation, and the AoA and range estimation MSE using the enhanced CSI estimation. The global parameter setting is listed as follows.

The carrier frequency is set to 28 GHz, the subcarrier interval is $\Delta {f} =$ 480 kHz, the antenna interval, $d_a$, is half the wavelength, the array sizes of the user and the BS are $1\times 1$ and $8 \times 1$, respectively, the number of paths is $L = 2$, and the reflection factor for the NLoS path, ${\beta _{C,1}}$, follows $\mathcal{CN}(0, 1)$. The numbers of neurons for hidden layers of CNNs are $C_1 = C_2 =$ 4. The number of OFDM symbols for each packet is $P_s = $ 14. 
The variance of the Gaussian noise is $\sigma_n^2 = 4.9177\times10^{-12} $ W. According to \eqref{equ:channel_vec} and \eqref{equ:receive_vec}, the SNR of each antenna element of BS can be expressed as 
\begin{equation}\label{equ:gamma_c}
	{\gamma _c} = {{P_t\sum\limits_{l = 0}^{L - 1} {{{\left| {b_{C,l}} \right|}^2}} } \mathord{\left/ 
			{\vphantom {{P_t^U\sum\limits_{l = 0}^{L - 1} {{{\left| {{b_{C,l}}\chi _{TX,l}^U} \right|}^2}} } {\sigma _N^2}}} \right.
			\kern-\nulldelimiterspace} {\sigma _n^2}}.
\end{equation}

The transmit power is determined according to the given SNR and  $\sigma_n^2$. Then, we first introduce the procedures for generating training data for the ISAC complex CNN CSI enhancer. 

\subsection{Generation of Training, Evaluation, and Test Data}

Training and evaluation CSI data are used to train the ISAC complex CNN and evaluate its generalization performance in the training process, respectively. Moreover, the test CSI data is used to obtain the final performance of the trained ISAC complex CNN.

Overall, we adopt two sorts of channels, namely, the static and dynamic channels, to test the performance of the ISAC complex CNN CSI enhancer. The static channel is suitable for the situation where UE and scatterers are all static, and the true CSI of the static channel does not change. On the contrary, the parameters of the dynamic channel are time-varying.
Then, we introduce the procedures for generating the static and dynamic channel CSI data, respectively.

\begin{figure}[!t]
	\centering
	\includegraphics[width=0.35\textheight]{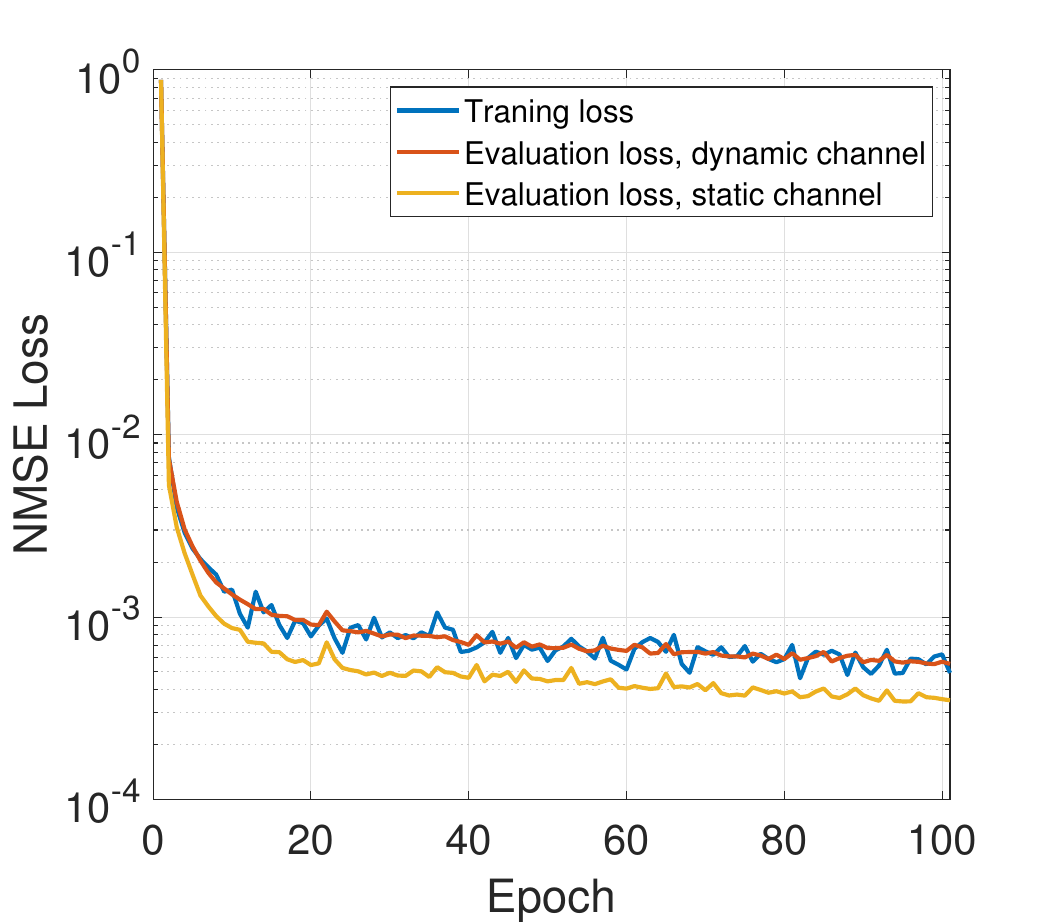}%
	\DeclareGraphicsExtensions.
	\caption{The NMSE loss in the training process of the ISAC complex CNN on the static and dynamic channel CSIs.}
	\label{fig:training_loss}
\end{figure}

\subsubsection{Static Channel CSI Generation}
The AoAs for the LoS and NLoS paths are ${\theta _0} = {30^ \circ }$ and ${\theta _1} = {59.5^ \circ }$, respectively; the range between BS and UE is ${r_{0,1}} = 91.26$ m, the ranges between BS and scatterer, and between scatterer and UE are ${r_{1,1}} = 28.7$ m and ${r_{1,2}} = 71.6$ m, respectively; the relative velocities between BS and UE is ${v_{0,1}} = 0$ m/s; the velocities between BS and scatterer, and between scatterer and UE are ${v_{1,1}} = {v_{1,2}} = 0$ m/s since UE and scatterers are all static. Using the above parameters, we can generate the true CSI for the static channel according to Section~\ref{subsec:channel_model}.

For each given SNR, ${\gamma _c}$, we generate $M_s =$ 2000 items of true CSI, and the transmit power is determined according to \eqref{equ:gamma_c}. The initial estimated CSIs using the LS method according to \eqref{equ:LS_est} are stored as the training input data, and the corresponding true CSIs are stored as the targets for the training input. We set the SNR range to be $[0,1,...,15]$ dB. After generating all the training CSI data for all 16 SNRs, we shuffle the CSI data set and choose the first 3/4 part of it to be the training data set, and the rest to be the evaluation data set. The test data are generated using the above parameters independently, and we set $M_s =$ 1000 for generating the test data.

\subsubsection{Dynamic Channel CSI Generation}

The procedures of generating the training, evaluation, and test CSI data for the dynamic channel are mostly the same as those for the static channel, except that the range and velocities of the UE are set to be random for each packet. The range between UE and BS is uniformly distributed from 5 m to 150 m, and the relative velocity between UE and BS is uniformly distributed from -10 m/s to 10 m/s.

We can see that the static channel is a special case of the dynamic channel where the AoA and range parameters are fixed, and the velocities of UE and all scatterer targets are 0 m/s. Therefore, we use the CSI data of dynamic channels to train the ISAC complex CNN enhancer to avoid the possible overfitting problem caused by the monotonous data. Here, we set $N_c = 256$ to generate the training CSI data. After preparing all the data for the training of the ISAC complex CNN enhancer, we can construct and train the CNN as shown in Section~\ref{sec:ISAC_CNN_processing}.

\begin{figure}[!t]
	\centering
	\includegraphics[width=0.35\textheight]{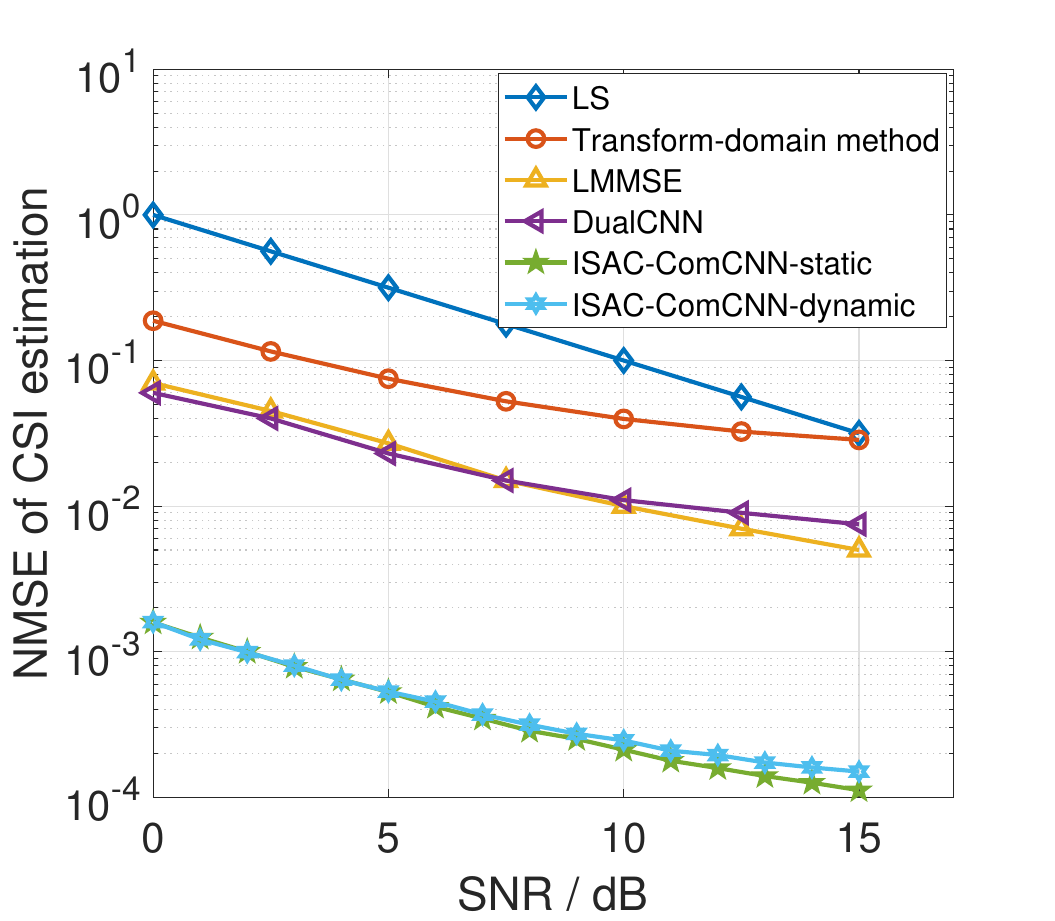}%
	\DeclareGraphicsExtensions.
	\caption{The NMSE of ISAC complex CNN compared with the existing methods.}
	\label{fig:NMSE_compare}
\end{figure}

\begin{figure}[!t]
	\centering
	\includegraphics[width=0.35\textheight]{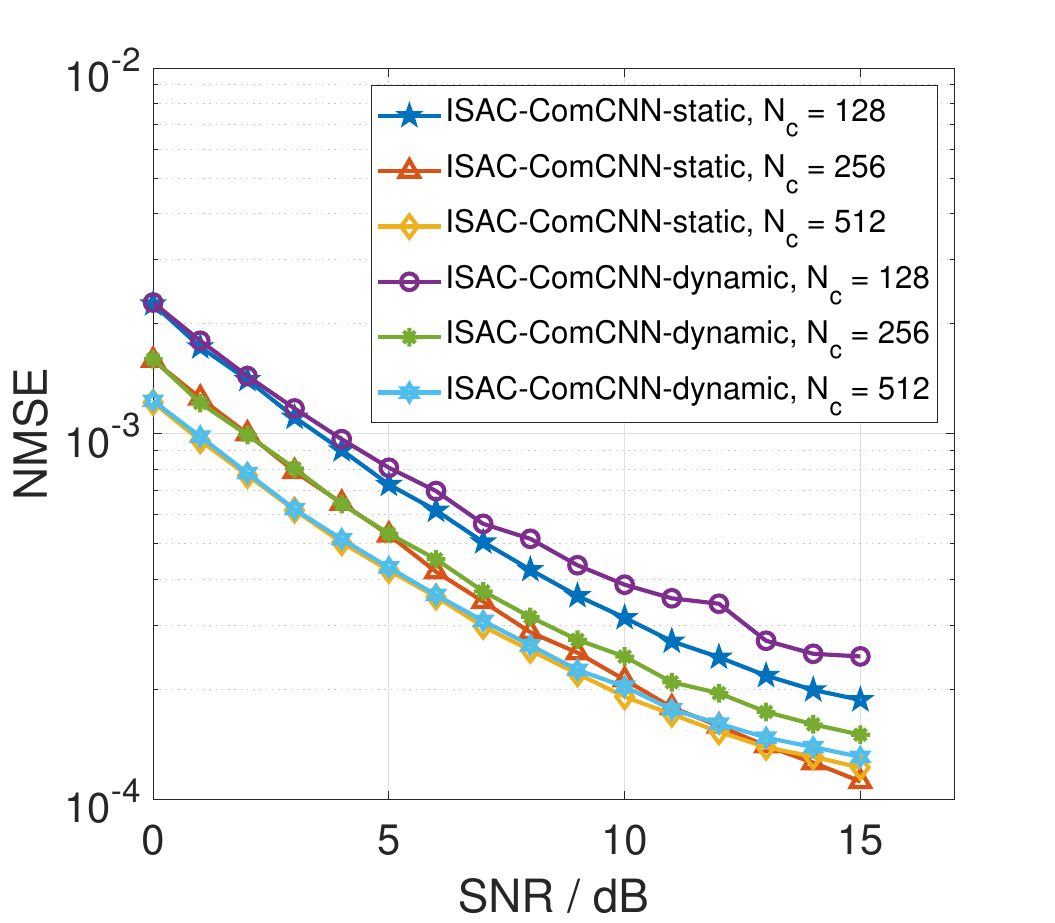}%
	\DeclareGraphicsExtensions.
	\caption{The NMSE of ISAC complex CNN in various numbers of $N_c$.}
	\label{fig:NMSE_ISAC}
\end{figure}

\subsection{Communication Performance using ISAC complex CNN}

In this subsection, we present the training and evaluation NMSE losses in the training process of ISAC complex CNN and the BER performance of communication demodulation using the CSI enhanced by the ISAC complex CNN.
The NMSE is calculated as
 \begin{equation}\label{equ:NMSE}
  \sigma _{\rm NMSE}^2 = \frac{{\left\| {{\bf{\hat H}} - {\bf{H}}} \right\|_F^2}}{{\left\| {\bf{H}} \right\|_F^2}},
 \end{equation}
 where ${{\bf{\hat H}}}$ is the CSI estimate, and ${\bf{H}}$ is the true CSI.
 
Fig.~\ref{fig:training_loss} presents the training and evaluation NMSE losses in the training process of the ISAC complex CNN. It can be seen that the loss of ISAC complex CNN decreases fast in the first 20 epochs, and hereafter it converges to a stable level within 30 epochs. We can see that the evaluation loss of ISAC complex CNN on the static channel CSI is slightly lower than that on the dynamic channel CSI. This is because the channel parameters of the dynamic channel are random, and there is thus a higher evaluation loss for the dynamic channel compared with the static channel. Overall, we can see that the evaluation loss is comparable to the training loss. This shows that the ISAC complex CNN has satisfactory generalization performance.

Fig.~\ref{fig:NMSE_compare} shows the CSI estimation NMSE of the ISAC complex CNN applied to the test CSI data. We choose the LS, LMMSE, DFT-based transform-domain, and the state-of-the-art dualCNN~\cite{2021JiangdualCNN} CSI estimators as comparisons, where dualCNN treats the real and imaginary parts of the CSI as two isolated real-value channels. The dualCNN shares similar complexity with the proposed ISAC complex CNN. The CSI estimation NMSEs for LMMSE and dualCNN are referenced to \cite{2021JiangdualCNN}. For the dualCNN trained based on the CSI data, it achieves comparable CSI estimation NMSE performance to the LMMSE method with lower complexity. We can see that the CSI estimation NMSE of the proposed ISAC complex CNN CSI enhancer is much lower than the above two CSI estimators by around 17 dB, given the same SNR. This is because the proposed ISAC complex CNN can maintain the complex-valued phase information without possible deviation caused by treating the real and imaginary parts of the complex CSI as isolated data, and the ISAC transform modules make the ISAC complex CNN able to process the sparse transform-domain CSI tensors more efficiently. The DFT-based transform-domain method also exploits the sparse CSI expression in the transform-domain by padding zeros to all the noise-like transform-domain CSI elements. However, it can not retain the phase of transform-domain CSI. Therefore, the CSI estimation NMSE of the DFT-based transform-domain method is significantly larger than that of our proposed enhancer. Moreover, the CSI estimation NMSE of ISAC complex CNN applied to the static channel is slightly lower than that applied to the dynamic CSI, since there is an inevitable regression error due to the randomness of the dynamic channel.

\begin{figure}[!t]
	\centering
	\includegraphics[width=0.35\textheight]{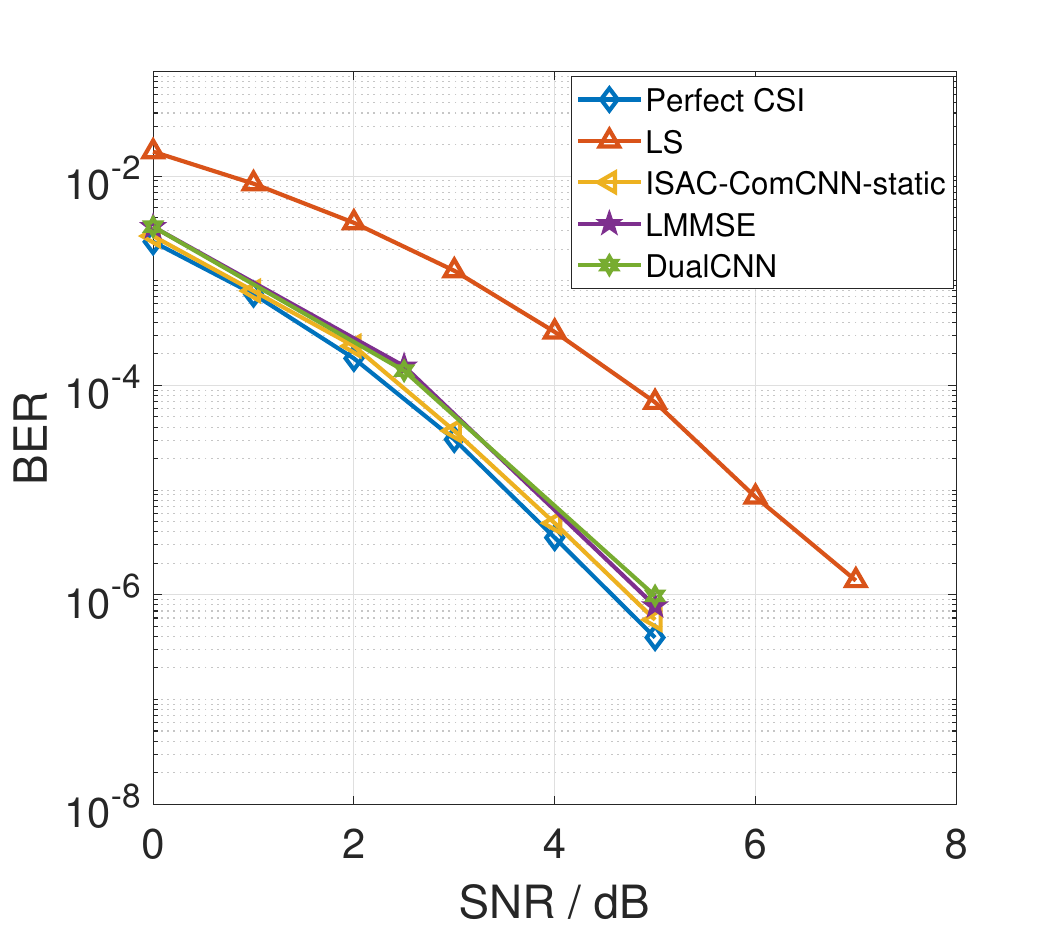}%
	\DeclareGraphicsExtensions.
	\caption{BER of communication demodulation using ISAC complex CNN enhanced static channel CSI compared with different CSI estimators under 4 QAM.}
	\label{fig:BER_4QAM_static}
\end{figure}

\begin{figure}[!t]
	\centering
	\includegraphics[width=0.35\textheight]{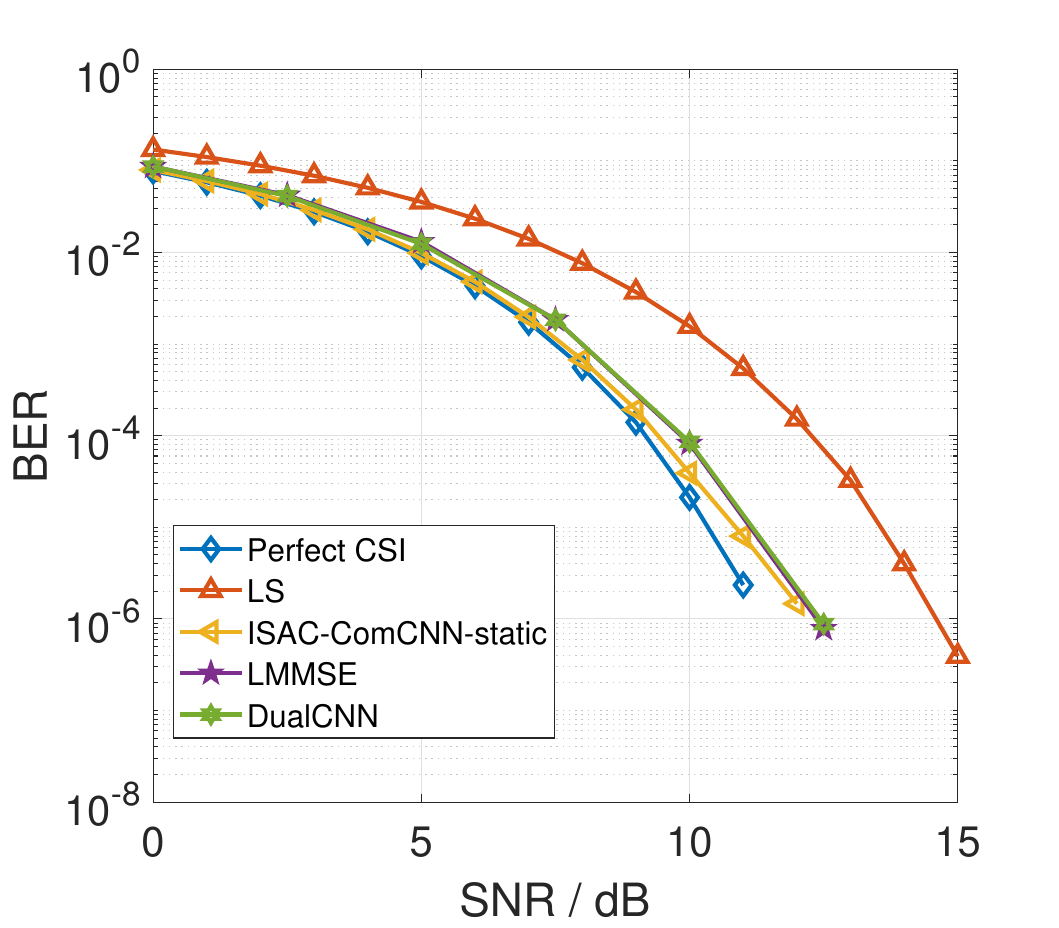}%
	\DeclareGraphicsExtensions.
	\caption{BER of communication demodulation using ISAC complex CNN enhanced static channel CSI compared with different CSI estimators under 16 QAM.}
	\label{fig:BER_16QAM_static}
\end{figure}

Fig.~\ref{fig:NMSE_ISAC} further shows the generalization performance of the ISAC complex CNN by using the complex CNN trained by dynamic channel CSI data with $N_c =$ 256 to enhance both the dynamic and static channel CSI data with various $N_c$.
We can see that for the same type of channel (dynamic or static channel), the higher the $N_c$ is, the lower the CSI estimation NMSE is. This means that the generalization performance of ISAC complex CNN increases as the dimensions of CSI increase. This is because the CSI data with higher dimensions generate radar heatmap images with higher resolution. Moreover, we can see that given the same $N_c$, the CSI estimation NMSE of the ISAC complex CNN applied to the static channel is slightly lower than that applied to the dynamic channel, which is consistent with the results in Fig.~\ref{fig:NMSE_compare}.

Then, we present the BER performance of communication demodulation using our proposed CSI enhancer compared with other CSI estimators. 

Fig.~\ref{fig:BER_4QAM_static} and Fig.~\ref{fig:BER_16QAM_static} show the BER performance of communication demodulation using the static channel CSI enhanced by the ISAC complex CNN under 4-QAM and 16-QAM modulation, respectively. It can be seen that when using 4-QAM, the BER performance using ISAC complex CNN enhanced CSI is similar to that using the perfect CSI estimation, which requires about 0.5 dB and 2.5 dB lower SNRs compared with the dualCNN and LS CSI estimators, respectively. When 16-QAM is used, ISAC complex CNN can achieve comparable BER performance to the perfect CSI estimation in the low SNR regime. In the high SNR regime, the BER using ISAC complex CNN enhanced CSI is slightly higher than that using the perfect CSI estimation but is still lower than those using dualCNN and LMMSE methods. This is because the CSI estimation NMSE of the proposed ISAC complex CNN is lower than the compared CSI estimators, as shown in Fig.~\ref{fig:NMSE_compare}. 

\begin{figure}[!t]
	\centering
	\includegraphics[width=0.35\textheight]{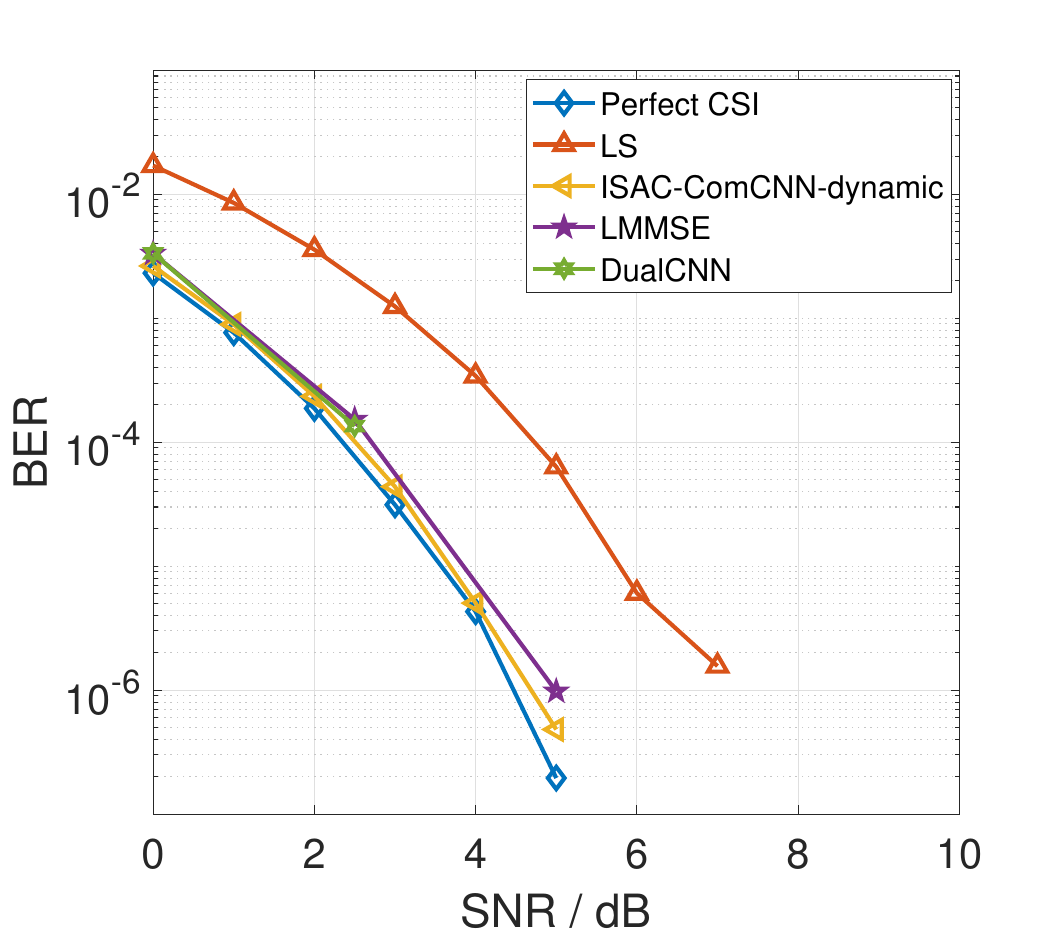}%
	\DeclareGraphicsExtensions.
	\caption{BER of communication demodulation using ISAC complex CNN enhanced dynamic channel CSI compared with different CSI estimators under 4 QAM.}
	\label{fig:BER_4QAM_dynamic}
\end{figure}

\begin{figure}[!t]
	\centering
	\includegraphics[width=0.33\textheight]{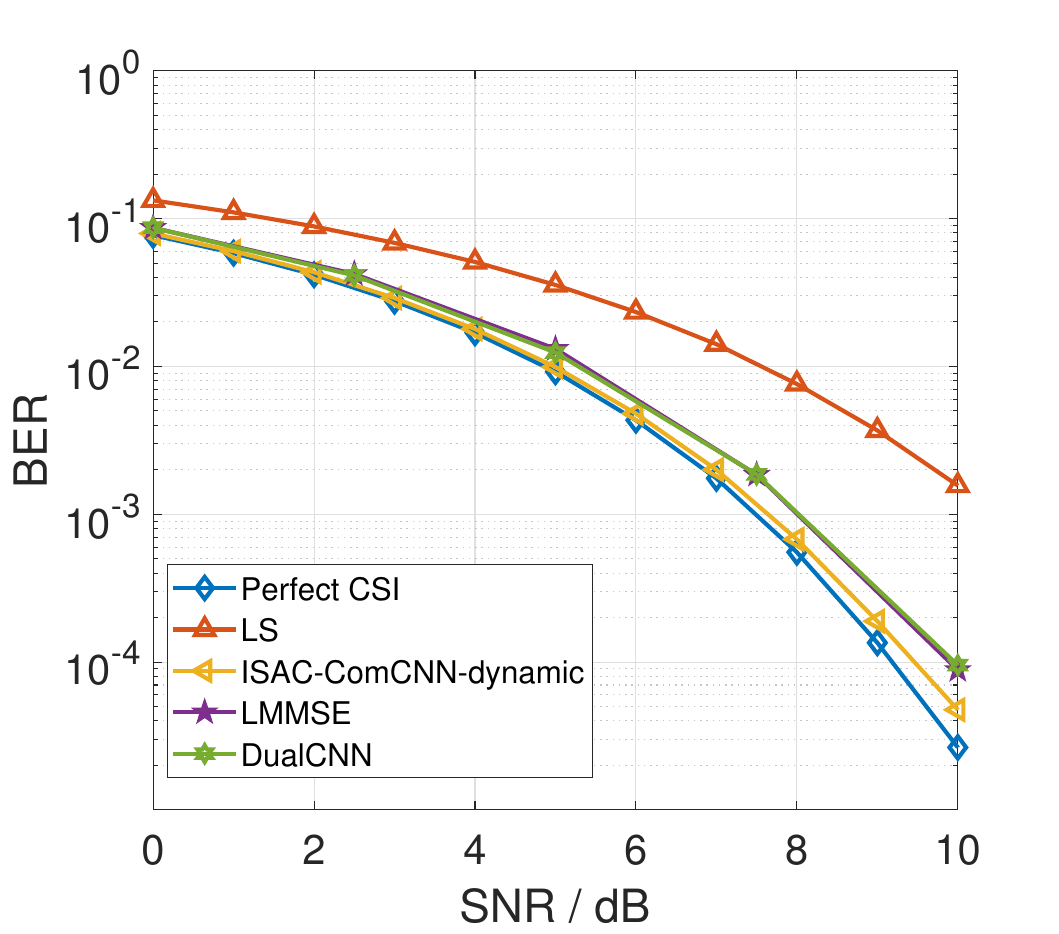}%
	\DeclareGraphicsExtensions.
	\caption{BER of communication demodulation using ISAC complex CNN enhanced dynamic channel CSI compared with different CSI estimators under 16 QAM.}
	\label{fig:BER_16QAM_dynamic}
\end{figure}

Fig.~\ref{fig:BER_4QAM_dynamic} and Fig.~\ref{fig:BER_16QAM_dynamic} show the BER of communication demodulation using the dynamic channel CSI enhanced by ISAC complex CNN under 4-QAM and 16-QAM modulation, respectively. It can be seen that when using 4-QAM, the BER using ISAC complex CNN is slightly higher than that with perfect CSI estimation, which requires about 0.3 dB and 2.3 dB lower SNRs compared with the dualCNN and LS CSI estimators, respectively. When 16-QAM is used, ISAC complex CNN can achieve comparable BER performance to the perfect CSI estimation in the low SNR regime. In the high SNR regime, the BER using ISAC complex CNN is slightly higher than that using the perfect CSI estimation but is lower than those using dualCNN and LMMSE methods. 

Overall, based on the above simulation results, we can see that the ISAC complex CNN can improve the accuracy of CSI estimation and BER performance for both static and dynamic channel situations.

\begin{figure}[!t]
	\centering
	\includegraphics[width=0.35\textheight]{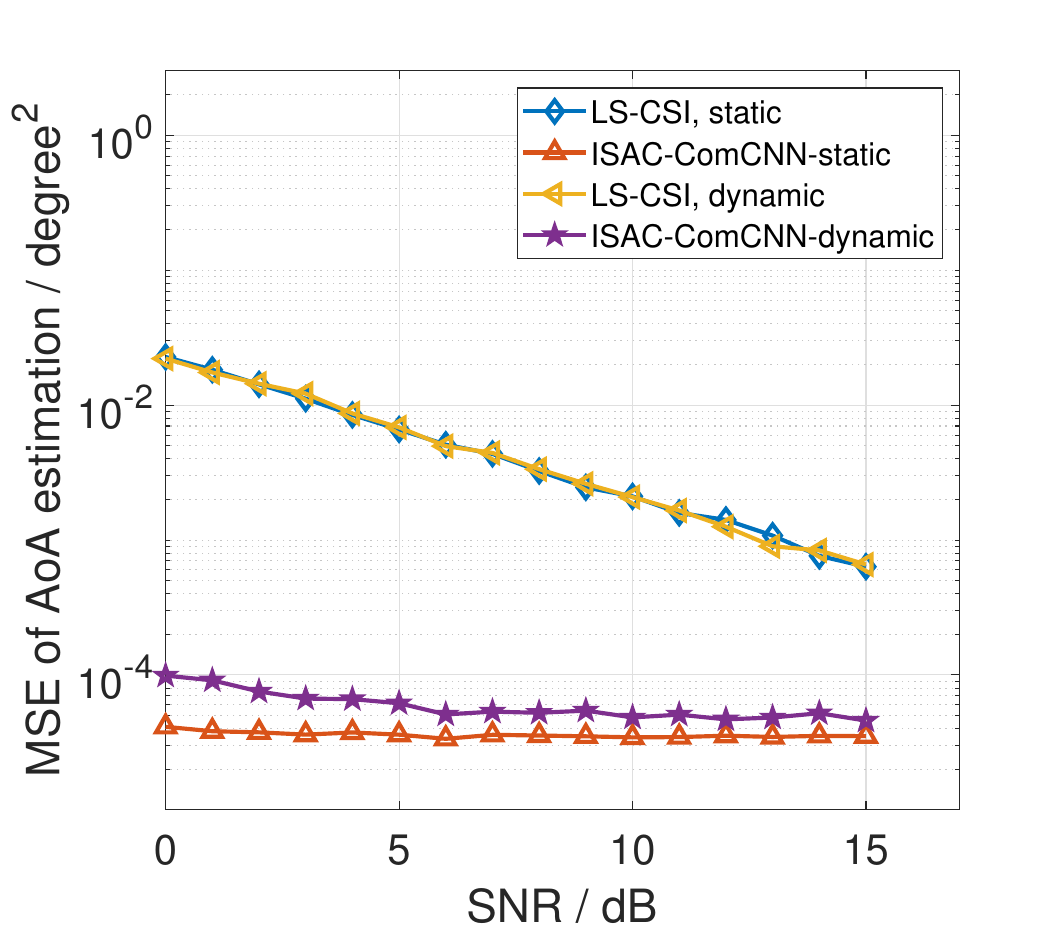}%
	\DeclareGraphicsExtensions.
	\caption{AoA estimation MSE based on the ISAC complex CNN enhanced CSI compared with that based on the conventional LS-estimated CSI, when $N_c =$ 256.}
	\label{fig:AoA_MSE}
\end{figure}

\subsection{Sensing Performance} \label{sec:sensing_performance}

Fig.~\ref{fig:AoA_MSE} illustrates the AoA estimation MSE based on the ISAC complex CNN enhanced CSI using the AoA estimation method shown in Section~\ref{sec:AoA_est}, and we consider both the situations of static and dynamic channel CSIs. It is shown that the MSE of AoA estimation based on the ISAC complex CNN enhanced CSI is significantly lower than those with the LS estimated CSI by more than 17 dB for both the static and dynamic channels. This is because the CSI estimation NMSE of ISAC complex CNN is significantly lower than that of LS methods. Moreover, we can see that the MSE of AoA estimation based on the static channel CSI enhanced by the ISAC complex CNN is slightly lower than that based on the enhanced dynamic channel CSI. This is because the randomness of the dynamic channel CSI leads to a larger estimation NMSE compared with the enhanced static channel CSI.

Then, we show the range estimation MSEs of the proposed biased FFT-based sensing method compared with the existing FFT-based and subspace-based sensing methods~\cite{Sturm2011}.

\begin{figure}[!t]
	\centering
	\includegraphics[width=0.35\textheight]{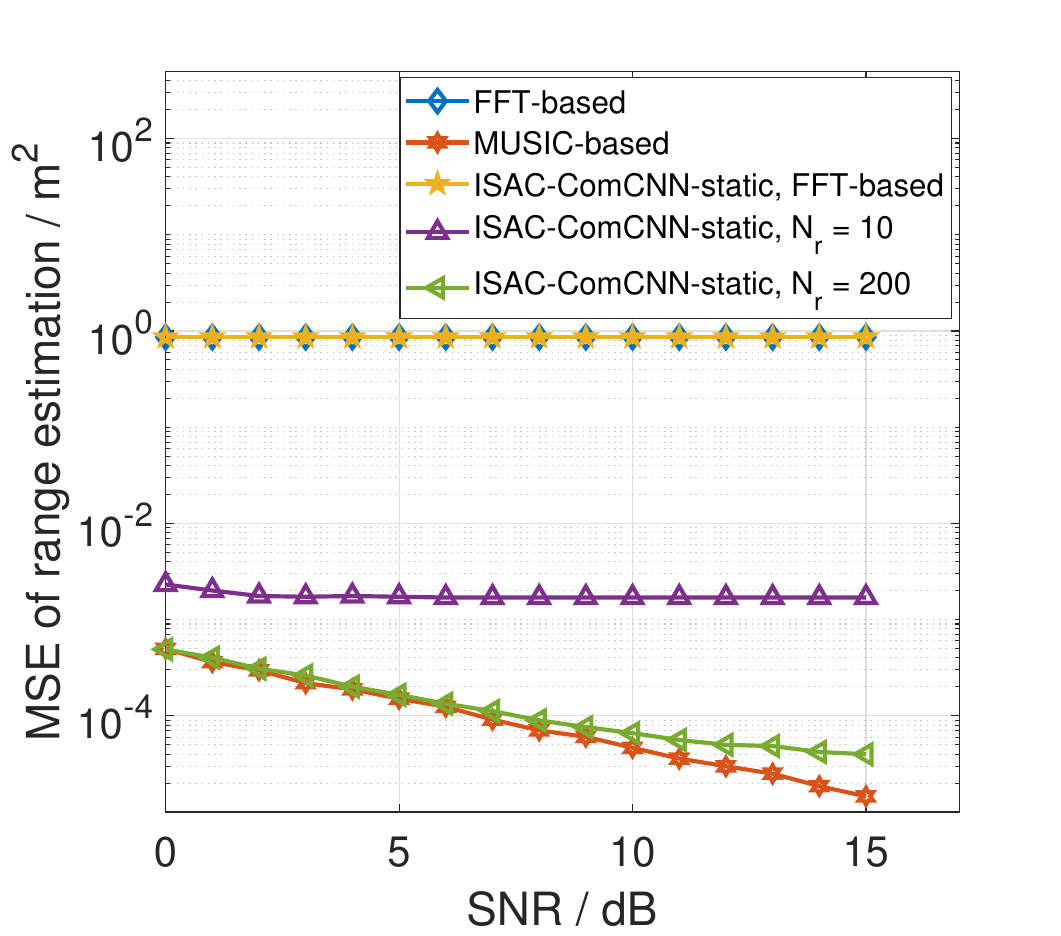}%
	\DeclareGraphicsExtensions.
	\caption{The range estimation MSE of the biased FFT-based sensing method using the static channel CSI under different $N_r$.}
	\label{fig:Range_MSE_static}
\end{figure}

\begin{figure}[!t]
	\centering
	\includegraphics[width=0.35\textheight]{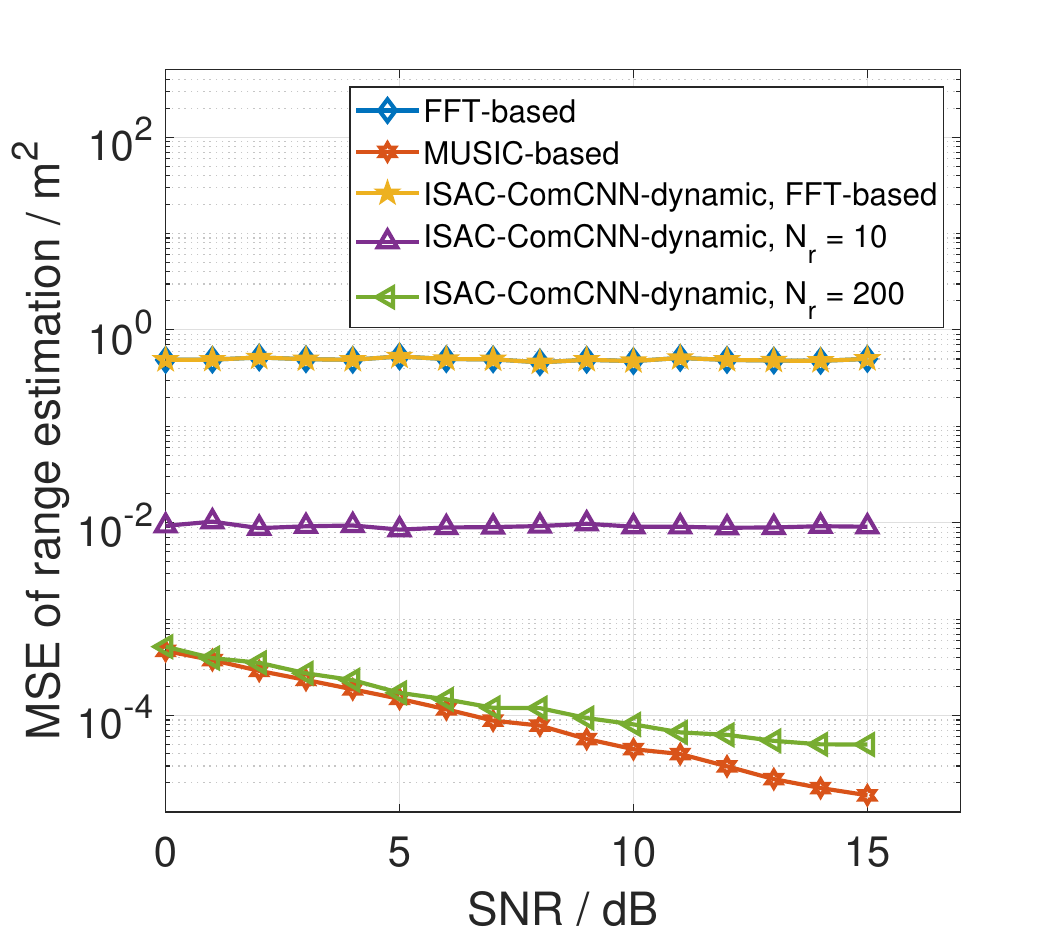}%
	\DeclareGraphicsExtensions.
	\caption{The range estimation MSE of the biased FFT-based sensing method using the dynamic channel under different $N_r$.}
	\label{fig:Range_MSE_dynamic}
\end{figure}

Fig.~\ref{fig:Range_MSE_static} and Fig.~\ref{fig:Range_MSE_dynamic} present the range estimation MSEs of the proposed biased FFT-based sensing method based on the static and dynamic channel CSIs, respectively. From Fig.~\ref{fig:Range_MSE_static}, we can see that the range estimation MSE of the proposed biased FFT-based sensing method is significantly lower than those of the FFT-based sensing method. This is because the average of the biased sensing results is the weighted sum that is closer to the true value as proved in \textbf{Appendix \ref{appendix:proof}}, while the MSE of the conventional FFT-based sensing method is restricted by the resolution, i.e., the grid interval of the range spectrum. Moreover, it is shown that the range estimation MSE of the biased FFT-based sensing method decreases with the increase of $N_r$ given the same SNR. This is because as $N_r$ increases, the bias terms become denser, and the latter term in \eqref{equ:r_est_bias} is thereby more accurate. 
By comparing Fig.~\ref{fig:Range_MSE_static} with Fig.~\ref{fig:Range_MSE_dynamic}, we can see that the range estimation MSE of the biased FFT-based sensing method applied to the static channel CSI is lower than that applied to the dynamic channel CSI. This is because the CSI estimation NMSE for the dynamic channel is slightly larger than that of the static CSI, as shown in Fig.~\ref{fig:NMSE_compare}, which leads to a larger error for range sensing based on the dynamic CSI estimation. When $N_r =$ 200, the range estimation MSE of the biased FFT-based sensing method is approximate to that of the high-complexity MUSIC-based sensing method in the low SNR regime both for the static and dynamic channel CSIs. The gap between the range estimation MSEs of the biased FFT-based and MUSIC-based sensing methods is enlarged with the increase of SNR in the high SNR regime. This is because the sensing MSE for each bias term in the biased FFT-based sensing method is relatively larger than those for the noise-subspace vectors in the MUSIC-based method.

\section{Conclusion}\label{sec:conclusion}
In this paper, we propose an ISAC complex CNN CSI enhancer that uses the complex-valued computation layers to form the CNN structure to maintain the phase information of complex-valued CSIs. Within the CNN CSI enhancer, the FFT-based ISAC transform modules are integrated into the complex CNN structure to transform the CSI data from the original domain into the sparse delay-angle-domain. The sparse delay-angle-domain CSIs can be regarded as heatmap images and can be efficiently processed by CNN, which improves the loss descending speed and generalization performance of the ISAC complex CNN. The MSE of AoA estimation using the enhanced CSI is improved significantly compared with that using the original CSI estimation. Finally, we propose a novel biased FFT-based sensing method, which significantly improves the range sensing accuracy at a similar complexity to the conventional FFT-based sensing method. Extensive simulation results show that the ISAC complex CNN CSI enhancer can converge within 30 training epochs, and the NMSE of its CSI estimates is about 17 dB lower, compared to the-state-of-art comparison estimators. Moreover, the BER of communication demodulation using the enhanced CSI estimates approaches that with the perfect CSI, and the mean square errors (MSEs) of AoA estimation based on the enhanced CSIs are significantly lower than those based on the original CSI estimates. Furthermore, the range estimation MSEs of the proposed biased FFT-based sensing method approaches those of the subspace-based sensing method, at a much lower complexity.


\begin{appendices} 
	\section{Estimation of $\hat L$}
	\label{appendix:L_est}
	Exploiting ${{\bf{v}}_\Sigma }$, we obtain a differential vector, ${{\bf{v}}_\Delta } \in {\mathbb{R}^{\left( {P - 1} \right) \times 1}}$, where ${\left[ {{{\bf{v}}_\Delta }} \right]_i} = {\left[ {{{\bf{v}}_\Sigma }} \right]_i} - {\left[ {{{\bf{v}}_\Sigma }} \right]_{i + 1}}$. According to \eqref{equ:eigenvalue}, we can see that ${\left[ {{{\bf{v}}_\Delta }} \right]_i} \approx 0$ for $i \ge L$, while $\left| {{{\left[ {{{\bf{v}}_\Delta }} \right]}_i}} \right| \gg 0$ for $i < L$. Based on this property, we calculate the mean value of the latter part of ${{\bf{v}}_\Sigma }$, and obtain
	\begin{equation}\label{equ:v_mean}
		\bar v = \frac{{\sum\limits_{k = \left\lfloor {\left( {P - 1} \right)/2} \right\rfloor }^{P - 1} {{{\left[ {{{\bf{v}}_\Delta }} \right]}_k}} }}{{P - \left\lfloor {\left( {P - 1} \right)/2} \right\rfloor }},
	\end{equation}
	where $\bar v$ should be an extremely small positive value that can be used to decide whether there is a signal. Based on the maximum likelihood criterion, the estimate of $L$ is obtained as
	\begin{equation}\label{equ:L_est}
		\hat L = \mathop {\arg \max }\limits_i {\left[ {{{\bf{v}}_\Delta }} \right]_i} > \left( {1 + \varepsilon } \right)\bar v,
	\end{equation}
	where $\varepsilon $ is a parameter to avoid the error caused by small noise. In this paper, we set $\varepsilon = 0.5$.
	
	\section{Proof of $\left| {{{\hat r}_{est}} - {r_t}} \right| < \left| {{r_{est}} - {r_t}} \right|$}
	\label{appendix:proof}
	
	We first consider the prerequisite condition that ${r_1} < {r_t} \le \frac{{{r_1} + {r_2}}}{2}$, and there is ${r_{est}} = {r_1} = \mathop {\arg \min }\limits_{r \in \left\{ {{r_1},{r_2}} \right\}} \left| {{r_t} - r} \right|$. Furthermore, we have
	\begin{equation}\label{equ:abs_1}
		\left| {{r_{est}} - {r_t}} \right| = {r_t} - {r_1},
	\end{equation}
	\begin{equation}\label{equ:abs_2}
		\bar k > \frac{1}{{{r_\delta }}}\left( {\frac{{{r_1} + {r_2}}}{2} - {r_t}} \right),
	\end{equation}
	
	Then, we continue the proof in two conditions, i.e., ${\hat r_{est}} \le {r_t}$, and ${\hat r_{est}} > {r_t}$. 
	
	1) When ${\hat r_{est}} \le {r_t}$, we have $\left| {{{\hat r}_{est}} - {r_t}} \right| = {r_t} - {\hat r_{est}}$. Therefore, we need to prove ${r_t} - {\hat r_{est}} < {r_t} - {r_1}$, i.e., ${\hat r_{est}} - {r_1}> 0$. According to \eqref{equ:r_est_bias}, ${\hat r_{est}} - {r_1} = {r_\delta }\left[ {\frac{{{N_r}\left( {{N_r}/2 - \bar k} \right)}}{{{N_r} + 1}}} \right] > 0$ is satisfied. 
	
	2) When ${\hat r_{est}} > {r_t}$, we need to prove ${\hat r_{est}} - {r_t} < {r_t} - {r_1}$, i.e., ${r_\delta }\left[ {\frac{{{N_r}\left( {{N_r}/2 - \bar k} \right)}}{{{N_r} + 1}}} \right] < 2\left( {{r_t} - {r_1}} \right)$. According to \eqref{equ:abs_2}, we have $\frac{{{N_r}}}{2} - \bar k < \frac{{{N_r}}}{2} + \frac{{2{r_t} - \left( {{r_1} + {r_2}} \right)}}{{2{r_\delta }}}$. By multiplying $\frac{{{r_\delta }{N_r}}}{{{N_r} + 1}}$ to the above inequality, we can obtain
	\begin{equation}\label{equ:abs_3}
		\begin{array}{l}
			{r_\delta }\left[ {\frac{{{N_r}\left( {{N_r}/2 - \bar k} \right)}}{{{N_r} + 1}}} \right] < \frac{{{N_r}}}{{{N_r} + 1}}\left( {\frac{{{r_\delta }{N_r}}}{2} + \frac{{2{r_t} - \left( {{r_1} + {r_2}} \right)}}{2}} \right)\\
			= \frac{{{N_r}}}{{{N_r} + 1}}\left( {\frac{{\left( {{r_2} - {r_1}} \right) + 2{r_t} - \left( {{r_1} + {r_2}} \right)}}{2}} \right)\\
			= \frac{{{N_r}}}{{{N_r} + 1}}\left( {{r_t} - {r_1}} \right) \le 2\left( {{r_t} - {r_1}} \right).
		\end{array}
	\end{equation}
	
	Therefore, the proof is completed for the prerequisite condition, ${r_1} < {r_t} \le \frac{{{r_1} + {r_2}}}{2}$. When ${r_2} > {r_t} > \frac{{{r_1} + {r_2}}}{2}$, the proof is similar to the above procedure, and we omit it in this paper.

\end{appendices}


%

{\small
	\bibliographystyle{IEEEtran}
	\bibliography{reference}
}
\vspace{-10 mm}
\ifCLASSOPTIONcaptionsoff
  \newpage
\fi

\end{document}